\documentclass[fleqn,usenatbib]{mnras}

\usepackage[T1]{fontenc}
\DeclareRobustCommand{\VAN}[3]{#2}
\let\VANthebibliography\thebibliography
\def\thebibliography{\DeclareRobustCommand{\VAN}[3]{##3}\VANthebibliography}

\usepackage{comment}
\usepackage{xcolor}
\usepackage{widetext}
\usepackage{ae,aecompl}
\usepackage{pdflscape}

\usepackage{graphicx}	% Including figure files
\usepackage{amsmath}	% Advanced maths commands
\usepackage{multicol} 
\usepackage{setspace}
\usepackage[center]{titlesec}
\usepackage{bm}

\usepackage[export]{adjustbox}
\usepackage[varg]{newtxmath}
\everymath{\displaystyle}
\def \mpc {\mbox{\rm Mpc}}

\title[MG-MAMPOSSt for modified gravity]{MG-MAMPOSSt: A code to test modifications of gravity with internal kinematics and lensing analyses of galaxy clusters}

\author[L. Pizzuti et al.]{
Lorenzo Pizzuti$^1$,\footnote{E-mail: pizzuti@oavda.it} 
Ippocratis D. Saltas$^2$
and Luca Amendola$^3$
\\
$^1$ Osservatorio Astronomico della Regione Autonoma Valle d'Aosta,  Loc. Lignan 39, I-11020, Nus, Italy\\
$^2$ CEICO, Institute of Physics of the Czech Academy of Sciences, Na Slovance 2, 182 21 Praha 8, Czechia\\
$^3$Institute of Theoretical Physics, Philosophenweg 16, Heidelberg University, 69120, Heidelberg, Germany\\
}

\date{Accepted XXX. Received YYY; in original form ZZZ}

\pubyear{2021}
\renewcommand{\thefootnote}{\fnsymbol{footnote}}

\begin{document}
\label{firstpage}
\pagerange{\pageref{firstpage}--\pageref{lastpage}}
\maketitle
\renewcommand{\thefootnote}{\arabic{footnote}}
\begin{abstract}
We present an upgraded version of \textsc{MG-MAMPOSSt}, an extension of the \textsc{MAMPOSSt} algorithm that performs Bayesian fits of models of mass and velocity anisotropy profiles to the distribution of tracers in projected phase space, to handle modified gravity models and constrain their parameters. The new version implements two distinct types of gravity modifications, namely general chameleon and Vainshtein screening, and is further equipped with a Monte-Carlo-Markov-Chain module for an efficient parameter space exploration. The program is complemented by the \textsc{ClusterGEN} code, capable of producing mock galaxy clusters under the assumption of spherical symmetry, dynamical equilibrium, and Gaussian local velocity distribution functions as in \textsc{MAMPOSSt}.
We demonstrate the potential of the method by analysing a set of synthetic, isolated spherically-symmetric dark matter haloes, focusing on the statistical degeneracies between model parameters. Assuming the availability of additional lensing-like information, we forecast the constraints on the modified gravity parameters for the two models presented, as expected from joint lensing+internal kinematics analyses, in view of upcoming galaxy cluster surveys. In Vainshtein screening, we forecast the weak lensing effect through the estimation of the full convergence-shear profile. For chameleon screening, we constrain the allowed region in the space of the two free parameters of the model, further focusing on the $f(\mathcal{R})$ subclass to obtain realistic bounds on the background field $|f_{\mathcal{R}0}|$. Our analysis demonstrates the complementarity of internal kinematics and lensing probes for constraining modified gravity theories, and how the bounds on Vainshtein-screened theories improve through the combination of the two probes.

\end{abstract}


%%%%%%%%%%%%%%%%%%%%%%%%%%%%%%%%%%%%%%%%%%%%%%%%%%

%%%%%%%%%%%%%%%%% BODY OF PAPER %%%%%%%%%%%%%%%%%%

\section{Introduction}
\makeatletter
\renewcommand{\@makefnmark}{\hbox{\textsuperscript{\tiny{\@thefnmark}}}}
\makeatother
Galaxy clusters constitute a very useful probe to test gravity theories, offering a rich laboratory to study the physics of dark matter, dark energy and gravity. Combinations of spectroscopic and lensing observations allow for the reconstruction of the cluster's mass profile, which in turn provide valuable information about the local gravitational potentials $\Phi$ and $\Psi$ that appear in the linearly perturbed metric. This idea has formed the basis for a multitude of phenomenological tests for dark energy theories beyond General Relativity (GR). The most popular theoretical frameworks in this context have been scalar-tensor theories, which extend GR through a new dynamical scalar field. Their theory space has seen strong constraints after the recent gravitational wave observations (e.g.  \citealt{Ezquiaga:2017ekz,Creminelli17,Sakstein:2017xjx,Baker:2017hug, Dima:2017pwp, Kobayashi:2018xvr, Amendola:2017orw}). The scalar field propagates a new fifth force which modifies the usual Newtonian potential at local scales leaving a phenomenological footprint on the internal kinematics and lensing of galaxy clusters. The family of scalar-tensor theories includes the conformally-coupled  theories such as Brans-Dicke or $f(\mathcal{R})$ gravity, or the more general Horndeski, beyond Horndeski and the recently introduced Degenerate Higher-Order Scalar-Tensor (DHOST hereafter) theories \citep{Zumalacarregui:2013pma,BenAchour:2016fzp,Langlois:2017mxy, Langlois:2018dxi, Kobayashi:2019hrl, Amendola:2019laa}. Previous works have placed strong constraints on $f(\mathcal{R})$ gravity through the abundance of galaxy clusters \citep{Schmidt2009, Rapetti2010,  Rapetti2011,  Ferraro2011, Lombriser:2010mp, Cataneo2015, Cataneo:2016iav}, or combining reconstructions of the cluster's thermal and lensing mass \citep{Terukina2012, Terukina:2013eqa, Wilcox:2015kna, Sakstein:2016ggl, Salzano:2016udu, Salzano:2017qac}. What is more, the combination of lensing and internal kinematics reconstruction of a galaxy cluster mass profile can provide a powerful test based on the gravitational slip parameter $\eta$, which defines the deviation from Einsteinian gravity. This was pursued by \citet{Pizzuti:2016ouw,Pizzuti:2019wte}, who forecasted galaxy cluster constraints on the gravitational slip following a simple phenomenological approach.

Building on previous works (eg.  \citealt{Pizzuti:2017diz}, hereafter Paper I), here we present the upgraded \textsc{MG-MAMPOSSt}, a new version of the \textsc{MAMPOSSt} code of   \citet{Mamon01} developed to determine galaxy clusters mass profiles by analysing the internal kinematics of member galaxies in the cluster. Given an input of projected positions of member galaxies and line of sight velocities, coming either from simulated or real clusters, the code solves the Jeans equation to reconstruct the local gravitational potential and the velocity anisotropy profile, under the assumptions of spherical symmetry and dynamical relaxation. A key output is a likelihood analysis for the free model parameters in the mass profile function from dynamics. In a modified gravity scenario, the gravitational potential receives a contribution from the additional degrees of freedom, producing an effective mass profile which differs from GR.
The first version of \textsc{MG-MAMPOSSt} has been presented in  Paper I for the case of linear $f(\mathcal{R})$ gravity, and has been applied to data of two galaxy clusters analysed within the CLASH/VLT collaborations (\citealt{Postman01,Rosati1}).
In the current version, we have extended the code's capability by implementing two popular and viable modifications of gravity: chameleon and beyond Horndeski gravity. These models represent two distinct screening mechanisms, namely the chameleon and Vainshtein ones, respectively. They are characterised by their own phenomenological impact on the inferred mass profile from kinematical data and require a specific study of their behaviour in this context. 

The \textsc{MG-MAMPOSSt} method is complemented by \textsc{ClusterGEN}, a Python3 code capable of generating synthetic galaxy clusters in a configuration of dynamical equilibrium, which can be used to perform tests of \textsc{MG-MAMPOSSt} and forecasting analyses neglecting all the systematic effects.

Our main goal is to showcase the code for the aforementioned class of modified gravity models, and present  an analysis of statistical degeneracy between the free parameters of the model, which will be essential when testing against real data. Therefore, we will  focus on a catalog of synthetic dark matter haloes produced by means of \textsc{ClusterGEN}. By assuming the availability of additional information on the galaxy cluster mass profiles, such as that provided by reliable lensing analyses, we also  forecast future constraints on the modified gravity parameters, through a proper implementation of the modified gravity dynamics in the Jeans equation. In a future paper, we will present an analysis of real galaxy cluster data. We parametrise the gravitational potential assuming   Navarro-Frenk-White model (NFW hereafter) of \citet{Navarro97} for the matter density perturbations. The NFW profile has been widely adopted and it has been shown to provide adequate fit for simulated DM haloes and for real clusters' data, both in GR ( e.g. refs. \citealt{Biviano01,Umetsu16,Peirani17}) and in modified gravity (e.g. \citealt{Lomb12,Wilcox:2016guw}). \\

The paper is structured as follows. 
In Section \ref{sec:theo} we present a brief summary about the two modified gravity models of this work. In Section \ref{sec:Codes} we present the \textsc{MG-MAMPOSSt} code, its features and the new modules implemented with respect to the original version of   \citet{Mamon01}. We further introduce the main properties of \textsc{ClusterGEN}, the generator of synthetic haloes accompanying \textsc{MG-MAMPOSSt}. Sections \ref{sec:beyond} and \ref{sec:fr} are dedicated to the applications of \textsc{MG-MAMPOSSt} over a set of simulated dark matter haloes in the case of beyond Horndeski and chameleon gravity respectively. In Section \ref{sec:Summary} we discuss our results and main conclusions. 

\begin{table}
\begin{center} 
\begin{tabular}{ |c|c|c| } 
\hline
{\bf Screening type}  & {\bf Models} & {\bf Key eqn's} \\
\hline
{\rm Chameleon} & $f(R)$, Brans-Dicke & (\ref{eq:cham})\\
\hline
{\rm Vainshtein} & Beyond Horndeski, DHOST &   (\ref{Poisson-BH}), (\ref{ippo-BH-Psi}) \\ 
\hline
\end{tabular}
\caption{The two distinct families of models considered in this work labelled according to their screening mechanism. Notice that, the linear Horndeski model described in the text is a very special case of the chameleon models where the chameleon mechanism is switched off, leading to the standard Yukawa force. For more details, we refer to the discussion of Section \ref{sec:theo}.  }
\label{table:models}
\end{center}
\end{table}

\section{Chameleon and Vainshtein screened theories}\label{sec:theo}

The vast majority of scalar-tensor theories rely on a screening mechanism to recover agreement with local gravity tests at small scales.  The dominant screening mechanisms are the so--called chameleon and Vainshtein mechanism (see e.g.  \citealt{Koyama:2015oma}), and the current version of the \textsc{MG-MAMPOSSt} code implements families of theories associated with both mechanisms.

In the presence of a screening mechanism, the fifth force that modifies the gravitational interaction w.r.t. GR is allowed to operate on sufficiently large scales, but gets suppressed locally. Both chameleon and Vainshtein mechanisms rely on the non-linear interactions of the scalar field to weaken the coupling between the scalar field and matter. However, each of them operates differently: whereas chameleon screening needs the non-linear potential interactions of the scalar, Vainshtein relies on higher-order derivative self-interactions of the scalar field. Luckily, neither one does not completely eliminate the trace of the fifth force, but leaves a characteristic imprint at local scales. Typical examples of theories exhibiting chameleon screening are conformally- coupled models such as $f(\mathcal{R})$ models or scalar-tensor models of the Brans-Dicke type. On the other hand, Vainshtein screening is a natural feature of the more general interactions in Horndeski, beyond Horndeski and DHOST scalar-tensor theories. The (derivative) interactions present in the latter theories generalise the simpler conformally-coupled interactions. In the following, in order to simplify the language, we choose to classify the models based on their screening mechanism, and we will refer to the conformally-coupled theories just as {\bf ``chameleon screening (CS) models"}, and to those exhibiting more general interactions as {\bf ``Vainshtein screening (VS) models"}. Indeed, the character of the underlying interactions is closely related to the associated type of screening operating in the theory. The {\bf ``linear Horndeski''} models which we will discuss later, correspond to a special case of conformally-coupled models where the chameleon screening is switched off, hence leading to a standard (unsuppressed) Yukawa force mediated by the scalar. An overview of the models considered in our work can be found in Table \ref{table:models}.

The different phenomenology of CS and VS mechanisms is due to the structurally distinct way the resulting fifth force depends on the source's density or its gradients. In theories with chameleon screening, the fifth force is sourced by the scalar field  $\phi$ ( e.g. \citealt{Khoury13}),
\begin{equation}\label{eq:cham}
\nabla^2 \phi = \frac{\partial V}{\partial \phi} + \mathcal{Q}\frac{ \rho}{M_\text{P}} e^{\mathcal{Q} \phi/(M_\text{P}c^2_l)}, 
\end{equation}
where $M_\text{P}=(8\pi G)^{-1/2}$ is the reduced Planck mass, $G$ is the Newton's constant and $c_\text{L}$ is the speed of light. $\mathcal{Q}$ is a dimensionless coupling constant\footnote[1]{In the literature the coupling constant is often indicated by $\beta$. Throughout this paper we use $\mathcal{Q}$ instead to avoid confusion with the anisotropy profile parameter (see Section \ref{sec:Codes}).}, $\rho$ the density of the matter field and $V(\phi)$ is the potential of the scalar field and is a  monotonic function of $\phi$. A typical form of the potential assumes a power-law $V(\phi) =\Lambda^{4+n}\phi^{-n}$ where $n$ and $\Lambda$ are constants. The quantity $\phi/M_{\text{P}}$ has the dimension of energy per unit mass.  The above equation captures a wide range of theories depending on the value of the coupling $\mathcal{Q}$. In particular, $ \mathcal{Q} = 1/\sqrt{6}$ for $f(\mathcal{R})$ gravity. 

Deep within the massive source, the scalar field is everywhere close to a minimum value $\phi_{s}$ and field gradients are negligible $\nabla^2 \phi \approx 0$ (see e.g. \citealt{Khoury2004} for details). Thus, from eq. \eqref{eq:cham} we obtain the solution of the scalar field inside the source:
\begin{equation}\label{eq:inner}
\phi_\text{int} \approx \left(\mathcal{Q}\frac{ \rho}{n\Lambda^{4+n}M_\text{P}} \right)^{-1/(n+1)},
\end{equation}
where we have used the power-law expression for the scalar potential and we have assumed  $\mathcal{Q} \phi/(M_\text{P}c^2_l) \ll 1$. 
In this regime, the fifth-force is screened.

Towards the outskirts of the source, the gradient of the field grows, and leads to a fifth-force per unit mass given by 
\begin{align}
F_{\phi} = -\frac{\mathcal{Q} }{M_\text{P}} \frac{\text{d}\phi}{\text{d}r}, \label{eq:force}
\end{align}
such that the gradient of the gravitational potential is modified as 
\begin{equation} \label{eq:pot_chameleon}
\frac
{\text{d}\Phi}{\text{d}r}=\frac{GM(r)}{r^2}+\frac{\mathcal{Q}}{M_\text{P}}\frac{\text{d}\phi}{\text{d}r}.
\end{equation}
where $M(r)$ is the mass of the object enclosed within radius $r$. Note that, above and in the following, we compute the gradient of the gravitational potential assuming spherical symmetry.
Far away from the massive body, the Laplacian term dominates over the field potential (i.e. $\partial V(\phi)/\partial \phi \ll \nabla^2\phi$) and the equation of motion for the exterior chameleon field is given by:
\begin{equation}\label{eq:outer}
\nabla^2 \phi_\text{ext} \approx \mathcal{Q}\frac{ \rho}{M_\text{P}}\,.
\end{equation}
For $f(\mathcal{R})$ gravity, the additional degree of freedom is expressed in terms of the scalaron field (e.g. \citealt{1988PhLB..214..515B}) 
\begin{equation}
|f_\mathcal{R}|\equiv\left|\frac{\text{d} f}{\text{d} \mathcal{R}}\right| = \exp\left[-\sqrt{\frac{2}{3}} \frac{\phi}{M_\text{P}c^2_l}\right]-1. \label{eq:frcham}
\end{equation} 
Equations \eqref{eq:inner} and \eqref{eq:outer} require the input of a profile for the matter density $\rho$ in order to be solved. In this work, we will model the halo mass density assuming a NFW profile \citep{Navarro97} and, following e.g.  \citet{Terukina2012}, we will use an analytical approximation for the chameleon field radial profile $\phi(r)$ matching the inner and outer solutions (see Section \ref{sec:Codes}). Given such an expression for $\phi(r)$, one can use Eq. (\ref{eq:force}) to derive the force acting on a test particle.

For beyond Horndeski and DHOST theories, which agree with gravitational wave observations, it has been shown that the Vainshtein screening mechanism is partly broken within massive sources. This leads to a different fifth force term, which does not depend only on the source's density, but also on its gradient  \citep{Kobayashi:2014ida,Crisostomi:2017lbg,Dima:2017pwp}:
\begin{equation}
\frac{\text{d} \Phi}{\text{d}r} =  \frac{G M(r)}{r^2} +  \frac{Y_{1} G}{4}  \frac{\text{d}^{2} M}{\text{d}r^2}. \label{Poisson-BH} 
\end{equation}

Here, $Y_1$ is the dimensionless coupling of the new force. The gradient of the mass profile is  $\text{d}M(r)/\text{d}r = 4 \pi \rho(r) r^2$, where $\rho$ is the total mass density. 
In a galaxy cluster, the density $\rho(r)$ is assumed to be the total matter density.  Expressing the second derivative of the mass in terms of matter density we find:
\begin{equation}\label{eq:BH_Mamon}
\frac{\text{d} \Phi}{\text{d}r} =  \frac{G M(r)}{r^2} \left[1+\frac{3}{4}Y_1\left(\frac{\rho(r)}{\bar{\rho}(r)}\right)\left(2+\frac{\text{d}\ln \rho}{\text{d}\ln r}\right)\right].
\end{equation} where $\bar{\rho}(r)$ is the average density at radius $r$.
Notice that, outside the mass distribution, the fifth-force effect vanishes because $\rho(r)\ll \bar\rho(r)$. What is more, the fifth force becomes zero at the scale radius $r_\text{s}$ where the logarithmic slope of $\rho(r)$ equals $-2$, regardless of the choice of the mass profile.  For a given acceleration, a value $Y_1 > 0$ leads to a  lower mass  with respect to GR in the region $r<r_\text{s}$, while it increases the mass  for $r>r_\text{s}$. The opposite happens for $Y_1<0$.

Eq. \eqref{eq:BH_Mamon} can be expressed in a Poisson GR-like form by defining an effective dynamical mass such that
\begin{equation} \label{eq:BH_effectiveMass}
  \frac{\text{d}\Phi}{\text{d}r} \equiv \frac{GM_{\text{dyn}}}{r^2}=\frac{G}{r^2}(M+M_1),   
\end{equation}
where  
\begin{equation}
  M_1= Y_1\pi r^3 \left(2\rho+r\frac{\text{d}\rho}{\text{d}r}\right).
\end{equation}
Currently, the most stringent upper and lower bounds are of the order $|Y_1| < 10^{-2} - 10^{-1}$ and come from astrophysical probes
(\citealt{Sakstein:2013pda,Sakstein:2015aac,Sakstein:2015zoa, Jain:2015edg,Babichev:2016jom,Dima:2017pwp,Saltas:2018mxc,Sakstein:2018fwz,Creminelli:2018xsv,Babichev:2018rfj,Ishak:2018his,Saltas:2019ius, Crisostomi:2019yfo}), while, on cosmological scales, constraints on $Y_1$ obtained with galaxy clusters are only of order unity (e.g. \citealt{Sakstein:2016ggl}). Note that the analyses on small and large scales are based on totally different physics, thus offer two independent and complementary ways to test the universality of the coupling constant in these models.\\ 

A fundamental difference between the two previous theories is that, whereas conformally-coupled theories such as $f(\mathcal{R})$ leave lensing unaffected, this is no longer true within beyond Horndeski and DHOST theories. In the latter case, the relativistic potential $\Psi$ is also modified explicitly as 
\begin{align}
\frac{\text{d} \Psi}{\text{d}r} =  \frac{G M(r)}{r^2} -  \frac{5 Y_{2} G}{4 r}  \frac{\text{d} M}{\text{d}r}, \label{ippo-BH-Psi}
\end{align}
where $Y_2$ is a dimensionless coupling constant. As we did for Eq. \eqref{Poisson-BH}, we can rewrite Eq. \eqref{ippo-BH-Psi} highlighting the modified gravity contribution in the following way:
\begin{equation}
    \frac{\text{d} \Psi}{\text{d}r} =\frac{G M(r)}{r^2}\left[1-\frac{15}{4}Y_2\left(\frac{\rho}{\bar{\rho}}\right)\right]. 
\end{equation}
Lensing is sourced by the combination
\begin{equation}
\Phi_{\text{lens}} = \frac{1}{2}(\Phi + \Psi)\label{eq:phi-lens},
\end{equation}
and it may provide constraints on the additional coupling $Y_2$ present in these theories (e.g.  \citealt{Sakstein:2018fwz}). In particular, the lensing potential $\Phi_{\text{lens}}$ obeys a GR-like Poisson equation $\nabla^2\Phi_{\text{lens}}=4\pi G \rho_\text{lens}$, with
\begin{equation}
    \rho_\text{lens}=\left(1+\frac{3}{4}Y_1-\frac{15}{8}Y_2\right)\rho+\left(\frac{3}{4}Y_1-\frac{5}{8}Y_2 \right)r\frac{\text{d}\rho}{\text{d}r}+\frac{Y_1}{8}r^2\,\frac{\text{d}^2\rho}{\text{d}r^2}. \label{eq:BH_efflens}
\end{equation}
From the above equation, we can define an effective lensing mass which, assuming spherical symmetry, is given by
\begin{equation}\label{eq:lensmass}
M_{\text{lens}} =\frac{r^2}{2G}\left[\frac{\text{d}\Psi}{\text{d}r}+\frac{\text{d}\Phi}{\text{d}r}\right]=M+\pi r^3\left[\left(Y_1-\frac{5}{2}Y_2\right)\rho+\frac{Y_1}{2}r\frac{\text{d}\rho}{\text{d}r}\right],
\end{equation}
Note that $M_{\rm lens}$ is sensitive to both $Y_1$ and $Y_2$. This will be useful in analyses where dynamical probes are combined with lensing measurements. The effective lensing mass can be further expressed in terms of the dynamical mass as $M_\text{lens}=M_\text{dyn}+M_2$, where
\begin{equation}
 M_2=-\pi r^3\left[\left(Y_1+\frac{5}{2}Y_2\right)\rho+\frac{Y_1}{2}r\frac{\text{d}\rho}{\text{d}r}\right].  
\end{equation}
\section{The \textsc{MG-MAMPOSSt} and \textsc{ClusterGEN} codes} \label{sec:Codes}
Here we introduce the main features of \textsc{MG-MAMPOSSt}, focusing on the new modules implemented for modified gravity. We also briefly describe the \textsc{ClusterGEN} code, which is used to produce the mock clusters for the applications of \textsc{MG-MAMPOSSt} presented in Sections \ref{sec:beyond} and \ref{sec:fr}.
\begin{figure*}
\centering
\includegraphics[width=0.8\textwidth]{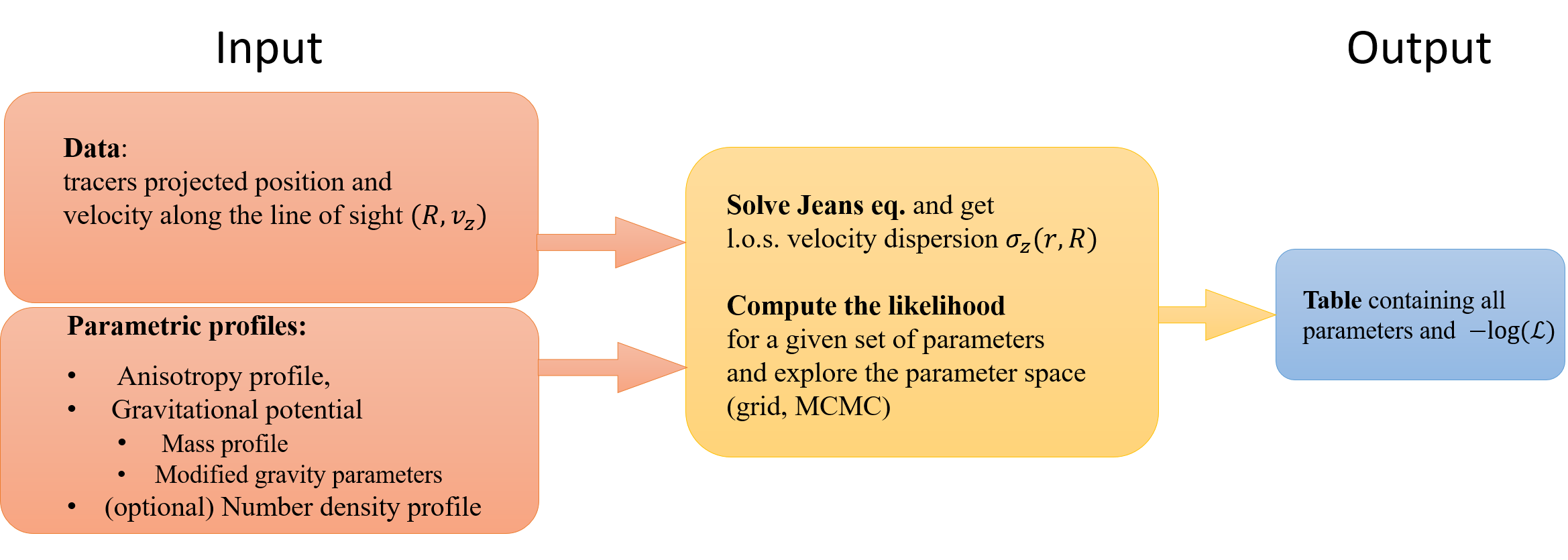}
\caption{\label{fig:sketch}Diagram which illustrates the structure of the \textsc{MG-MAMPOSSt} code.}
\end{figure*}

\subsection*{A. MG-MAMPOSSt}
\textsc{MG-MAMPOSSt} is a new version of the \textsc{MAMPOSSt} code of   \citet{Mamon01} for the dynamical analysis of collisionless  systems aimed at constraining modified gravity models. The method has been first applied in Paper I for the simple case of linear $f(\mathcal{R})$ gravity. The upgraded version of the code implements general non-linear chameleon screening and Vainshtein screening.
\\

{\bf Main concept:} The \textsc{MAMPOSSt} (Modelling Anisotropy and Mass Profile of Spherical Observed Systems) procedure, originally developed by \citet{Mamon01}, infers the mass profile and velocity anisotropy profile of galaxy clusters under the assumption of spherical symmetry and dynamical relaxation. The code uses the projected information on the cluster member galaxies, which are considered as collisionless tracers of the gravitational potential, i.e. the projected distances,  $R$, from the cluster center (hereafter, projected radii)  at which a galaxy is seen by the observer, and the line-of-sight (l.o.s.) velocities, computed in the rest frame of the cluster. Working in this projected phase space $(R,v_z)$, \textsc{MAMPOSSt} performs a Maximum Likelihood fit by solving the stationary spherical Jeans equation,
\begin{equation}\label{eq:ippo-jeans}
\frac{\text{d} (\nu \sigma_r^2)}{\text{d} r}+2\beta(r)\frac{\nu\sigma^2_r}{r}=-\nu(r)\frac{\text{d} \Phi}{\text{d} r}.
\end{equation}
In the above equation, $\nu(r)$ corresponds to the number density profile of tracers, $\sigma^2_r$ is the squared velocity dispersion along the radial direction, $\beta \equiv 1-(\sigma_{\theta}^2+\sigma^2_{\phi})/2\sigma^2_r$ is the velocity anisotropy profile and $\Phi$ is the total gravitational potential, which can provide a definition for the dynamical mass profile as
\begin{equation}
\frac{\text{d} \Phi}{\text{d}r} = \frac{GM_{\text{dyn}}(r)}{r^2}.
\end{equation} 
{\bf Anisotropy profile:} The anisotropy profile $\beta(r)$ is not known in principle and poses one of the major systematics in the dynamical mass-profile reconstruction. To infer it, both parametric (e.g.  \citealt{Mamon01,Brok13,Read21}) and  non-parametric methods have been developed. 
The first non-parametric determination of the velocity anisotropy profile was performed by \citet{Binney1982} and then improved by other works (e.g. \citealt{Mamon19} and references therein).  In \textsc{MG-MAMPOSSt}, we adopt a parametric approach with six different models as in the original version of \citet{Mamon01}. However, for illustrative purposes, we will rely only on the  anisotropy model of   \citet{Tiret01},
\begin{equation} \label{eq:tiret}
\beta_T(r)=\beta_\infty\frac{r}{1+r_\beta},
\end{equation}  
where $\beta_\infty$ is the velocity anisotropy for $r\to\infty$ and $r_\beta$ is the characteristic radius of $\beta_T(r)$ (anisotropy radius).
\\

{\bf Velocity and number density profiles:} The expression for the radial velocity dispersion can be obtained by integrating Eq. \eqref{eq:ippo-jeans} (e.g. \citealt{MamLok05})
\begin{equation}\label{eq:sigmajeans}
\sigma^2_r(r)=\frac{1}{\nu(r)}\int_r^{\infty}{\exp\left[2\int_r^s{\frac{\beta(t)}{t}\text{d}t}\right]\nu(s)\frac{\text{d}\Phi}{\text{d}s}\text{d}s},
\end{equation}
and the \emph{local} l.o.s. velocity dispersion is (\citealt{Binney1982})
\begin{equation}\label{eq:sigmaz}
\sigma^2_z(r,R)=\left[1-\beta(r)\left(\frac{R}{r}\right)^2\right],
\end{equation}
%\textcolor{red}{where the projected radius is given by $R=\sqrt{r^2-z^2}$.}
In our current version of the code we assume that the three-dimensional distribution of the velocity field is Gaussian, as all previous applications of \textsc{MAMPOSSt}. Note that there is no \emph{a priori} restriction on the choice of the 3D velocity profile. However, the tests performed by \citet{Mamon01} over a set of haloes from cosmological simulation showed that the Gaussian model works sufficiently well even
when the l.o.s. velocity distribution exhibits significant deviations from Gaussianity. The number density $\nu(r)$ can in general be measured directly from the phase space. As such, it can be excluded from the fit. We assumed that the tracer density profile $\nu(r)$ follows an NFW model. Since the normalization of the profile is factored out in Eq. \eqref{eq:sigmajeans} the model is entirely parametrised by the scale radius $r_{\nu}$ of the tracer. In general, the distribution of the cluster's members differs from the distribution of the total matter density (see e.g.  \citealt{Biviano06,budzynski12,Mamon19}). Therefore, the scale radii $r_\nu$ of the tracer density and $r_\text{s}$ of the total mass profile do not need to coincide. For additional information about the anisotropy models and mass profiles implemented in the original version, see  \citet{Mamon01}.
\\

{\bf Code output and sampling features:} \textsc{MG-MAMPOSSt} constrains the free  parameters for the mass and anisotropy profile, as well as for  the gravity theory under study. The output of the code is the logarithm of the total likelihood as
\begin{equation}
-\ln \mathcal{L}_{\text{dyn}}=-\sum_{i=1}^N \ln q(R_i,v_{z(i)}| \bf{\theta}).\label{eq:lnl}
\end{equation}
In Eq. (\ref{eq:lnl}), $N$ is the  number of member galaxies considered in the fit,  $q(R_i,v_{z(i)}| \bf{\theta})$ is the probability of observing a galaxy with projected position $R_i$ and l.o.s. velocity $v_{z(i)}$ given the model(s) described by the set of parameters $\bf\theta$, where

\begin{equation}
q(R,v_{z})=\frac{2\pi R \,g(R,v_{z})}{N_{\rm proj}(R_\text{max})-N_{\rm proj}(R_\text{min})},\label{q}
\end{equation}
where $N_{\rm proj}(R)$ is the predicted number of galaxies in the cylinder of  projected radius $R$, whose axis passes through the cluster center, $R_\text{max}$ and $R_\text{min}$ represent the maximum and minimum projected radii considered in the fit, respectively and $g(R,v_{z})$ is the surface density of observed objects with l.o.s. velocity $v_z$. Under the assumption of a 3D-Gaussian distribution for the velocity field, one has (\citealt{Mamon01}) 

\begin{equation}
    g(R,v_{z})=\sqrt{\frac{2}{\pi}}\int_{R}^{\infty}\frac{r\nu(r)}{\sigma_z(R,r)\sqrt{r^{2}-R^{2}}} \exp\left[-\frac{v_{z}^{2}}{2\sigma_{z}^{2}(R,r)}\right]\text{d}r,\label{eq:grv}
\end{equation}
with $\sigma_z(R,r)$ given by eq. \eqref{eq:sigmaz}.

In \textsc{MG-MAMPOSSt}, likelihoods are computed over a multi-dimensional grid of values in the parameter space. However, for a large number of parameters ($\gtrsim 4$) the grid is computationally too expensive when the code runs in Modified Gravity mode. In order to make our analysis more efficient, we have implemented a Monte Carlo-Markov Chain (MCMC) module, in which the chain is based on a simple Metropolis-Hastings algorithm with a fixed-step Gaussian random walk. This is particularly useful to perform forecasts over simulated catalogs with several haloes.
\footnote[2]{The public version of \textsc{MAMPOSSt} of \citet{Mamon01} (https://gitlab.com/gmamon/MAMPOSSt) is already equipped with an MCMC algorithm based on the CosmoMC package (\citealt{Lewis02}, https://cosmologist.info/cosmomc/). We are planning to include CosmoMC in our \textsc{MG-MAMPOSSt} version for future applications.}
Fig. \ref{fig:sketch} summarizes the structure of \textsc{MG-MAMPOSSt}.\\

{\bf Gravitational potential:} Compared to the original \textsc{MAMPOSSt} code, \textsc{MG-MAMPOSSt} introduces new parametrisations for the gravitational potential in Eq. \eqref{eq:sigmajeans}. All the models are based on a NFW mass density profile, but the next version of the code will extend this to other mass profiles. The gravitational potentials implemented in the current version are as follows:
\\

{\bf 1.} {\it Linear Horndeski models}: These models have been introduced in the first version of \textsc{MG-MAMPOSSt} and correspond to a subclass of Horndeski theories where the Newtonian potential $\Phi$ can be written in terms of three parameters, namely $h_1$, $Q^2$ and $m$ (see  Paper I and references therein).  These parameters are functions of redshift, but they can be considered constant at first approximation. 
In this class of models, there is no screening at all. In order to pass the local gravity constraints, either one assumes a very small coupling, or that the baryons are not sensitive to the fifth force. Here, we mention the Linearized Horndeski parameterisation for completeness. However, since the model has been already presented and tested in Paper I, we will not discuss this case any further in this paper.
\\

{\bf 2.} {\it Vainshtein Screening}: The form of the gravitational potential is according to Eq. \eqref{Poisson-BH}. The coupling constant $Y_1$ is the additional free parameter, along with the mass profile and velocity anisotropy parameters. We assume a NFW mass  profile to parametrise the density perturbations:
\begin{equation}\label{eq:NFWdens}
\rho(r)=\frac{\rho_\text{s}}{r/r_\text{s}(1+r/r_\text{s})^2},
\end{equation}
with $\rho_\text{s}$ the characteristic density and $r_\text{s}$ the scale radius at which the logarithmic derivative of the density profile equals $-2$. In this case, Eq. \eqref{eq:BH_Mamon} can be rewritten as
\begin{equation}  \label{BH-NFW}
\frac{\text{d}\Phi_\text{MG}}{\text{d}r}=\frac{GM_\text{NFW}(r)}{r^2}\left\{1+\frac{Y_1}{4}\frac{r^2(r_\text{s}-r)/(r_\text{s}+r)^3}{[\ln(1+r/r_\text{s})-r/(r_\text{s}+r)]}\right\},
\end{equation}
where
\begin{equation}
 M_{\text{NFW}}(r)=M_{200}\frac{\ln(1+r/r_\text{s}) -(r/r_\text{s})/(1+r/r_\text{s})}{\ln(1+c)- c/(1+c)}
\end{equation}
is the standard NFW mass profile in GR, and $c=r_{200}/r_\text{s}$ is the concentration. In the above equation $M_{200}=4\pi r_\text{s}^3\rho_\text{s}[\ln(1+c)-c/(1+c)]$ is the total mass of a sphere of radius $r_{200}$ enclosing an average matter density 200 times the critical density of the Universe at that redshift $\rho_\text{c}(z)=3H^2(z)/(8\pi G)$.

According to Eq. \eqref{eq:BH_effectiveMass}, the contribution induced by the modification of gravity is given by
%\begin{displaymath} 
%\frac{d\Phi}{dr}=\frac{GM_{200}}{r^2(\ln(1+c)- c/(1+c))}\left[\ln\left(1+\frac{r}{r_\text{s}}\right)\right.
%\end{displaymath}
%\begin{equation}
% \left.-\frac{r/r_\text{s}} {\left(1+r/r_\text{s}\right)}+\frac{Y_1}{4}\frac{r^2(r_\text{s}-r)}{(r_\text{s}+r)^3}\right], \label{BH-NFW}
%\end{equation}
\begin{equation}
    M_1(r)=M_{200}\frac{Y_1}{4}\frac{r^2(r_\text{s}-r)}{(r_\text{s}+r)^3}\times[\ln(1+c)- c/(1+c)]^{-1}.
\end{equation}

\begin{figure*}
\centering
\includegraphics[width=0.7\textwidth]{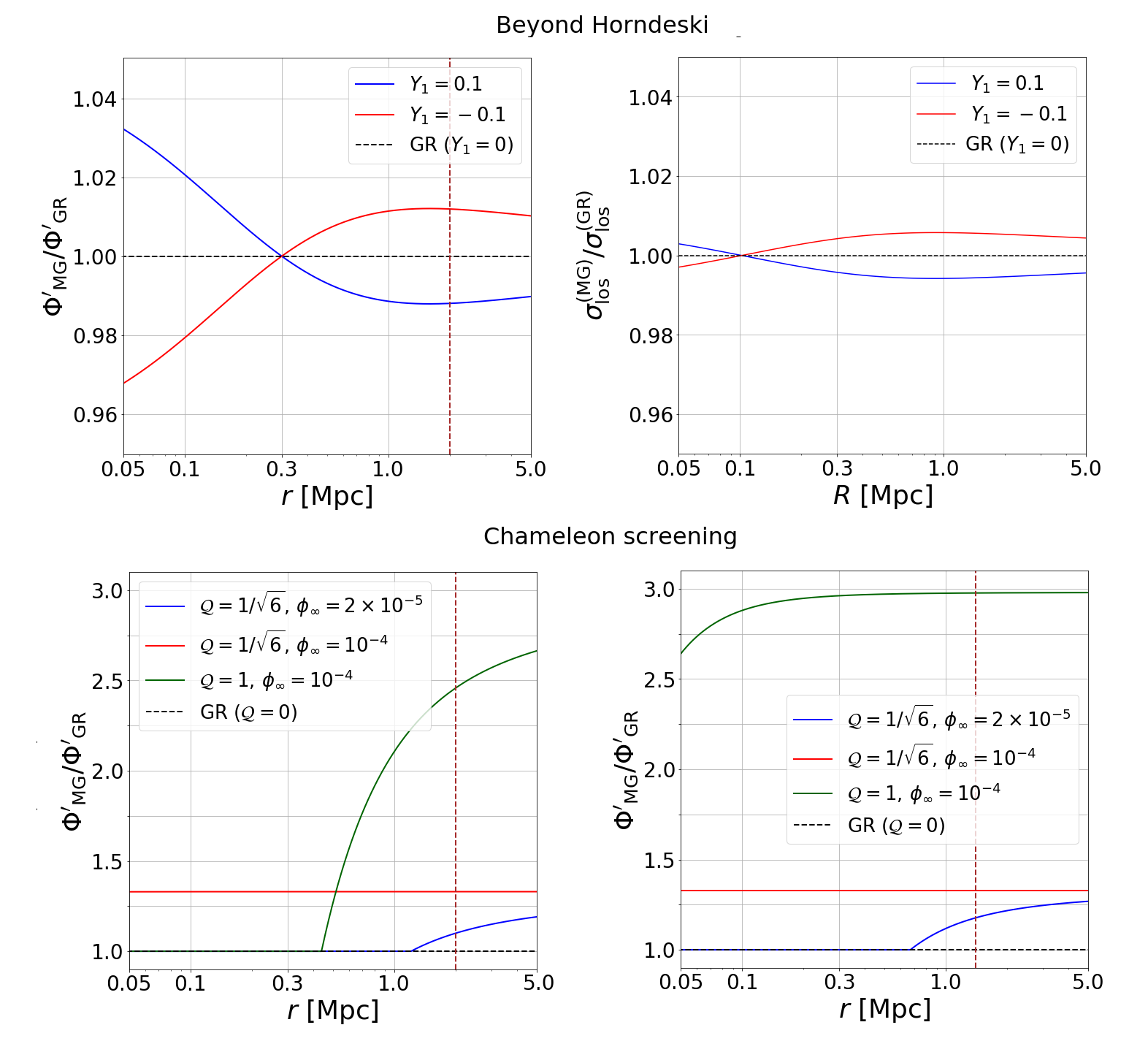}
\caption{\label{fig:BH_test} {\it Top left}: Ratios of the potential gradients as a function of the distance from the cluster center, for a NFW halo with $r_{200}=2.0\,\text{Mpc},\,\,r_\text{s}=0.3\,\text{Mpc}$.  The profiles $\Phi'_\text{MG}(r)$ are obtained in a VS model with $Y_1=0.1$ (red curves) and $Y_1=-0.1$ (blue curves), values compatible with current astrophysical constraints, while $\Phi'_\text{GR}$ represents the derivative of the Newtonian potential. {\it Top right}: Ratios of the corresponding velocity dispersion profiles $\sigma_{\text{los}}(R)$, integrated along the line of sight, at a projected position $R$.  {\it Bottom}: Ratios of the potential gradients in CS gravity with $\mathcal{Q}=1/\sqrt{6}$, $\phi_\infty=10^{-4}$ (red curves), $\mathcal{Q}=1/\sqrt{6},\,\,2\times10^{-5}$ (blue curves) and $\mathcal{Q}=1$, $\phi_\infty=10^{-4}$ (green curves). The left plot shows profiles derived from a NFW halo with $r_{200}=2.0\,\text{Mpc},\,\,r_\text{s}=0.3\,\text{Mpc}$ while the right plot is for haloes with $r_{200}=1.4\,\text{Mpc},\,\,r_\text{s}=0.33\,\text{Mpc}$. The vertical brown dashed lines in the left plots and in the bottom right plot represent the radius $r_{200}$.}
\end{figure*}

In the top left plot of Fig. \ref{fig:BH_test} we show the ratios of modified to Newtonian (GR) gravitational potentials.
%the ratios $\Phi'_{MG}(r)/\Phi'_{GR}(r)$, where $\Phi'_{MG}(r) \equiv \text{d}\Phi_{MG}/\text{d}r$ is the derivative of the gravitational potential in beyond Horndeski gravity and $\Phi'_{GR}$ is the same for the standard Newtonian potential. 
The profiles are computed adopting an NFW model with $r_{200}=2.0\,\text{Mpc},\,\,r_\text{s}=0.3\,\text{Mpc}$ for two different fifth-force coupling values, $Y_1=0.1$ (blue line) and $Y_1=-0.1$ (red line). These values are consistent with current bounds derived from astrophysical probes (e.g. \citealt{Sakstein:2018fwz,Creminelli:2018xsv, Saltas:2018mxc}).
In the case of $Y_1>0$ the potential gradient (i.e. the effective dynamical mass) is enhanced with respect to GR for radii smaller than $r_\text{s}$ and damped at large $r$. Exactly the opposite happens for $Y_1<0$. In the top right panel of \ref{fig:BH_test} we plot the corresponding ratios of l.o.s. velocity dispersions. The integration has been preformed by using eq. (A15) of \citet{MamLok05} where for the velocity anisotropy profile we have used the Tiret model with $\beta_\infty=0.5$ and $r_\beta=r_\text{s}$\footnote[3]{As for the number density profile, we still assume a NFW with $r_\nu\equiv r_\text{s}$}. 
For realistic values of the coupling parameter, the effect on cluster's internal kinematics is very small (order of $\sim 10^{-2}$). We thus expect that the constraints derived with the \textsc{MG-MAMPOSSt} procedure alone cannot reach the level of accuracy as those from astrophysical analyses.\\

{\bf 3.} {\it General Chameleon Screening}: For this class of models the gravitational potential is given by Eq. \eqref{eq:pot_chameleon}.
As mentioned in Section \ref{sec:theo}, for the chameleon field $\phi$ we use an analytic approximation that matches the two limiting solutions inside ($\nabla^2 \phi \approx 0$)  and outside the object ($\partial V/\partial \phi$ is subdominant). In the case of an NFW mass density profile, the field profile reads
\begin{equation}
\label{eq:field}
\phi(x) =
  \begin{cases}
%    \phi_\text{s}[r/r_\text{s}(1+r/r_\text{s})^2]\equiv \phi_\text{int}(r) &   r < S  \\
       \phi_\text{s}[x(1+x)^2]^{1/(n+1)}\equiv \phi_\text{int}(x) &   r < S  \\
     -\displaystyle{\frac{\mathcal{Q}\rho_\text{s}r_\text{s}^2}{M_\text{P}}}\displaystyle{\frac{\ln(1+x)}{x}}-\displaystyle{\frac{C}{x}}+\phi_{\infty}\equiv \phi_\text{ext}(x) & r > S. 
     %-\displaystyle{\frac{\mathcal{Q}\rho_\text{s}r_\text{s}^2}{M_\text{P}c^2_l}}\displaystyle{\frac{\ln(1+r/r_\text{s})}{r/r_\text{s}}}-\displaystyle{\frac{C}{r/r_\text{s}}}+\phi_{\infty}\equiv \phi_\text{ext}(r) & r > S.
  \end{cases}\
\end{equation}
In the above equation we have defined $x=r/r_\text{s}$, $\phi_{\infty}$ is the background value of the field,  $C$ is an integration constant and
\begin{displaymath}
\phi_\text{s} \equiv \left(\frac{n\Lambda^{n+4}M_\text{P}}{\mathcal{Q} \rho_\text{s}}\right)^{1/(n+1)}.
\end{displaymath}
The \emph{screening radius} $S$ represents the transition scale between the two regimes. $C$ and $S$ can be determined requiring that $\phi_ \text{int}$ and $\phi_\text{ext}$ and their derivatives match at $S$. 
In the halo's inner core the field is strongly suppressed with respect to its ambient value $\phi_\text{s} \ll \phi_{\infty}$, and we can approximate the interior solution to be negligible $\phi_\text{int} \simeq 0$. In this case, for $C$ and $S$, 
\begin{equation}
S=\frac{\mathcal{Q} \rho_\text{s}r_\text{s}^3}{M_\text{P}\phi_{\infty}}-r_\text{s},\label{eq:screenpar1}
\end{equation}
\begin{equation}
    C=-\frac{\mathcal{Q} \rho_\text{s}r_\text{s}^2}{M_\text{P}}\ln(1+S/r_\text{s})+\phi_{\infty}S/r_\text{s}. \label{eq:screenpar2}
\end{equation}

Chameleon $f(\mathcal{R})$ gravity is included as the sub-class obtained for $\mathcal{Q}=1/\sqrt{6}$.

 While for linear Horndeski and VS models the modified gravity parameters are independent of the size of the halo, in CS the suppression of the fifth force is related to the matter density. In particular, as we will show in Section \ref{sec:fr}, clusters with smaller mass and larger $r_\text{s}$ have a screening mechanism less efficient with respect to more massive haloes.

In the bottom plots of Fig. \ref{fig:BH_test}, we show the relative ratio of mass profiles in CS with respect to GR obtained for different values of $\mathcal{Q}$ and  $\phi_\infty$. For a large background field, the cluster is not screened and the effective mass profile is enhanced of a factor $2\mathcal{Q}^2$ compared to the standard NFW model. Decreasing $\phi_\infty$, the screening mechanism becomes active and $M_{\text{dyn}}\equiv M_{\text{NFW}}$ for $r\le S$. Grey curves show the effective mass profiles for a chameleon field $\phi_\infty=10^{-4}$ and a coupling constant $\mathcal{Q}=1$ -- since the screening effect is connected to the coupling with matter, part of the halo is screened, while the fifth force increases in the non-screened region. 
The bottom left plot of Fig. \ref{fig:BH_test} refers to more massive halo with higher concentration with respect to the bottom right plot (see details in the caption of the Figure). As expected, in this second case the screening suppression is reduced and the effect of the fifth force is enhanced also for small field values.

\subsection*{B. \textsc{ClusterGEN}}
The \textsc{ClusterGEN} code, first presented in   \citet{Pizzuti:2019wte} is a generator for isolated, self-gravitating systems in dynamical equilibrium, populated by collisionless point-like tracers according to a spherical model. The code produces mock-data catalogs which serve as input for \textsc{MG-MAMPOSSt}. The main purpose is to test the constraining power of the \textsc{MG-MAMPOSSt} method for different modified gravity scenarios in ideal conditions where all the systematics are under control. The synthetic clusters can be built with different characteristic masses and scales, while the internal kinematics is assumed to be governed by the Jeans equation \eqref{eq:ippo-jeans} with a 3D Gaussian local velocity distribution function (VDFs). Now dynamical systems
may have non-Gaussian local VDFs. The correct way to build a simulated spherical system in dynamical equilibrium is to use six-dimensional distribution functions, expressed in terms of the binding energy and angular momentum (\citealt{Kazantzidis04}), or in terms of action-angle variables (\citealt{vasiliev2019}). However, the \textsc{MAMPOSSt} procedure has been shown to work quite well in recovering the mass density and anisotropy profile of galaxy clusters (e.g. \citealt{Mamon01,Old15}) and dwarf spheroidal galaxies (\citealt{Read21}) assuming a Gaussian
local VDF. In this paper, we aim to test \textsc{MG-MAMPOSSt} in an ideal scenario where the mass profile reconstruction is unbiased with respect to the set of synthetic haloes. As such, our mock catalog, generated by assuming a 3D Gaussian local VDF, provides useful hints to study the degeneracy between the free parameters of the model in a modified gravity framework.

For a given radial number density profile, \textsc{ClusterGEN} distributes the particles in spherical shells from the cluster center and assigns to each tracer a velocity whose components, in spherical coordinates, have a squared dispersion $\bm{\sigma}^2_{\mathbf{r}} (r) = \{ \sigma^2_r(r), \sigma^2_{\theta}(r), \sigma^2_{\phi}(r)\}$, where $\sigma^2_r(r)$ is given by the integral solution Eq. \eqref{eq:sigmajeans},  while $ \sigma^2_{\theta}(r) = \sigma^2_{\phi}(r)=  [1-\beta(r)]\sigma^2_r$.

The current version of \textsc{ClusterGEN} works with four different anisotropy models, namely constant anisotropy $\beta(r)=\beta_\text{C}$ (which reduces to the isotropic case when $\beta_\text{C}=0$), the Mamon \& {\L}okas profile, the Tiret profile and the modified Tiret profile (see e.g.  \citealt{Biviano01}), which allows for negative anisotropy in the halo center.  As for the  gravitational potential, the radial velocity dispersion of Eq. \eqref{eq:sigmajeans} can be obtained both assuming GR as well as with one of the modified gravity models implemented in the \textsc{MG-MAMPOSSt} code. In this work, we only consider GR-based mock catalogues, leaving the analysis of non-GR mock haloes for future applications.
\begin{figure}
\centering
\includegraphics[width=0.434\textwidth]{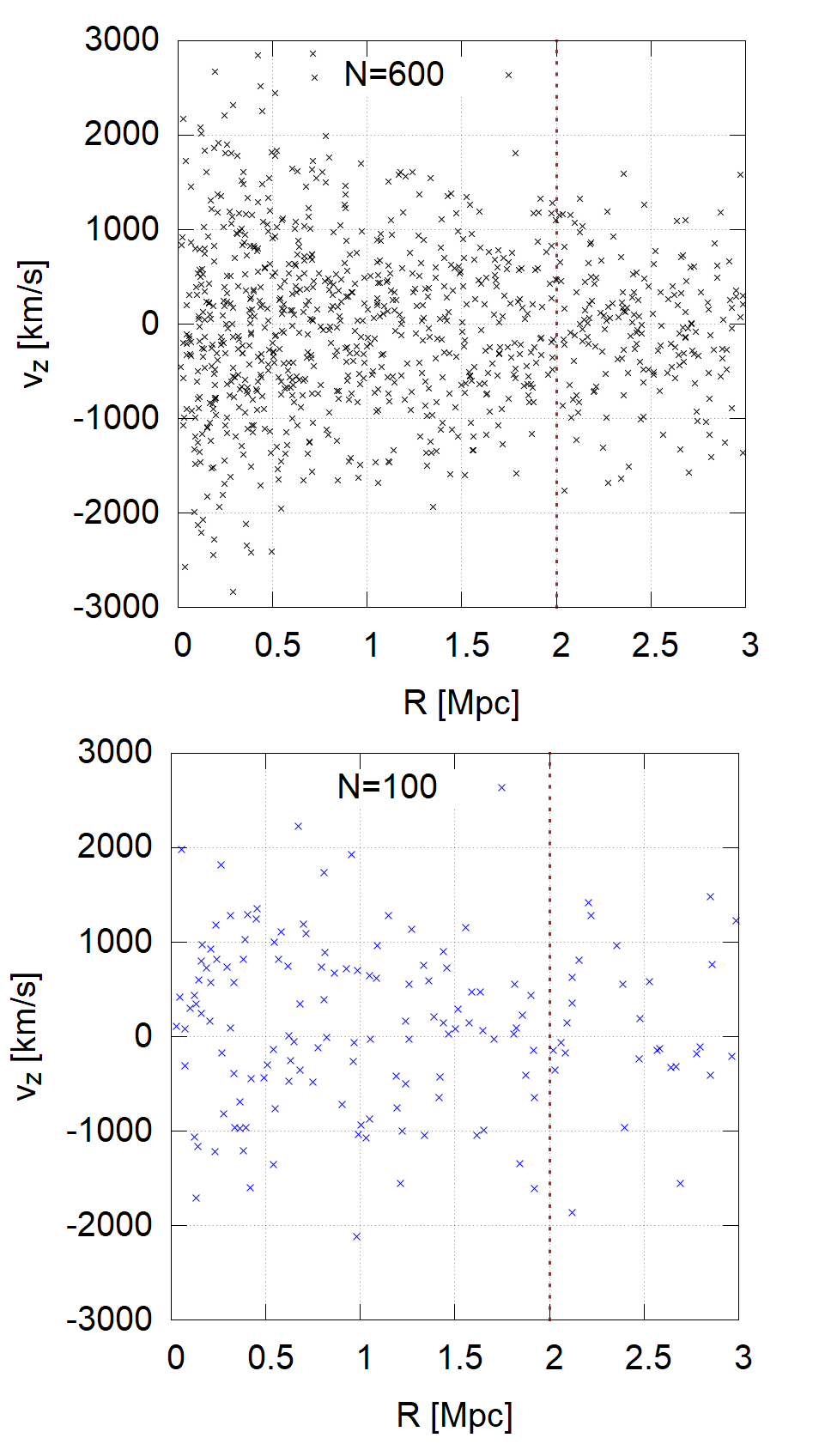}
\caption{\label{fig:pps}Projected phase spaces for a standard NFW halo with $r_{200}=2.0\,\text{Mpc},\,\,r_\text{s}=0.3\,\text{Mpc}$, assuming a Tiret model for the velocity anisotropy profile with $r_\beta=r_\text{s}$ and $\beta_\infty=0.5$. Up: 600 tracers considered within $r_{200}$ (vertical brown dashed line). Bottom: 100 tracers within $r_{200}$.}
\end{figure} 

The projected phase space $(R,\,v_z)$ used as input data for \textsc{MG-MAMPOSSt} is generated from the six-dimensional phase space using $R=(x^2+y^2)^{1/2}$. 
In Fig. \ref{fig:pps}, we plot two phase spaces obtained for one cluster in our catalog (see next Section for halo's parameters details) considering $N=600$ particles (left) and $N=100$ particles (right) within the virial cylinder ($R < r_{200}$).

\section{Application I: Vainshtein screening}
\label{sec:beyond}
In this Section, we show the application of \textsc{MG-MAMPOSSt} for the choice of VS, to spherically-symmetric and virialised synthetic clusters generated with the \textsc{ClusterGEN} code as outlined in Section \ref{sec:Codes}. We investigate the possible constraints on the additional degree of freedom $Y_1$ from internal kinematics analysis alone and simulating the availability of additional information such as the probability distribution provided by a strong+weak lensing probe.

\subsection{Setup}\label{sec:setup}
For VS models, modifications of gravity are not related to the size of the halo considered. Thus, all synthetic phase spaces are generated as different realization of the same massive NFW halo with $r_{200}=2.0\,\text{Mpc}$, $r_\text{s}=0.3\,\text{Mpc}$, and with a Tiret anisotropy profile $\beta(r)=\beta_{\infty}r/(r+r_\beta)$ with $r_\beta=r_\text{s}$ and $\beta_{\infty}=0.5$ (i.e. \citealt{MamLok05}). In view of the proof-of-concept aim of the paper, this choice has been made to facilitate the combined analysis of several clusters. Furthermore, in order to avoid spurious signatures of modified gravity due to statistical fluctuations in the cluster generation process, each phase space is derived requiring that the logarithm of the likelihood computed at the best fits of the standard \textsc{MAMPOSSt} analysis in GR differs from the one computed at the cluster true values by less than $ 1\%$. This way, the obtained constraints reflect the true statistical degeneracy between the model parameters. 

Note that real galaxy clusters at $z < 1$ generally exhibit lower values of $r_{200}$ and smaller concentration parameters (see e.g. \citealt{Biviano17}). For example, a typical galaxy cluster of $M_{200} \sim 10^{14.5} \,\text{M}_\odot$ at $z =0$ (with  $h=0.7$) has  $r_{200}=1.4\, \text{Mpc}$ and $r_\text{s}=0.33\,\text{Mpc}$ (i.e. $c\sim4.3$ according to e.g. \citealt{Dutton14}). However, we have calibrated our simulations to reflect the best fit values obtained by the \textsc{MAMPOSSt} analysis of the massive cluster MACS J1206.2-0847 (MACS 1206 hereafter) \citep{Biviano01} within the CLASH and CLASH-VLT collaborations. This has been done in view of the application of \textsc{MG-MAMPOSSt} over the MACS 1206 data-set in order to obtain realistic constraints on the MG models presented here. In the case of VS models, we will see that the values of $r_\text{s}$ and $r_{200}$ are not relevant for the analysis since the fifth force exhibits the same structure at all scales. As we have already mentioned in Section \ref{sec:Codes} this is not true for the CS mechanism. We will discuss in Section \ref{sec:fr} how our constraints change in this specific class of models when varying the values of the NFW parameters.

  As a default choice, each synthetic cluster is constructed with 600 particles (tracers) in the radial range $[ 0.05\,\mpc,r_{200} ]$. This choice is a reasonable, although optimistic, expectation for the number of member galaxies' spectroscopic redshifts available for few dozen clusters from ongoing and next generation surveys. 
It is, however, an important systematic factor in realistic observations -- indeed, Jeans analysis of galaxy clusters' internal kinematics is limited by several effects, such as substructures, modeling of velocity anisotropy profiles, departures from spherical symmetry and dynamical relaxation  (see e.g. \citealt{Pizzuti19b} and references therein). This may produce a bias in the mass profile reconstruction even for a quite large ($\sim 1000$) number of tracers, especially when searching for tiny deviations from standard gravity. Moreover, the new degrees of freedom could be strongly degenerate with the mass profile parameters and no constraints can be derived at all from internal kinematics (see again \citealt{Pizzuti19b} for the case of linear $f(\mathcal{R})$ gravity).

We will therefore discuss how our results are affected when lowering the number of tracers considered in the synthetic phase spaces as well as when considering stacked clusters. As concerns the projected radial range, we note that for real data the dynamics of member galaxies in clusters is dominated by the Brightest Central Galaxy (BCG) in the innermost region, which is thus excluded from the analysis. Similarly, even for relaxed clusters, departures from dynamical equilibrium can become important above $r_{200}$. The analysis of \citet{Falco13} showed that the Jeans equation can be still used to correctly infer the radial velocity dispersion up to $\simeq 1.4\,r_{200}$. However, since it is not clear what is the maximum projected radius $R$ below which Jeans analysis is valid, we make a conservative assumption of $R_\text{max}=r_{200}$.  For further discussions see e.g.  \citet{Biviano01,Pizzuti:2019wte}.

\subsection{Kinematical analysis}
We considered a catalog of 20 synthetic haloes generated in Newtonian gravity (GR). This is a fair expectation of the number of relaxed galaxy clusters for which very high quality imaging and spectroscopic data could be available from present and future surveys.
\subsubsection{Single cluster}
Given the tracers' phase space $(R,v_{z})$, we first apply \textsc{MG-MAMPOSSt} to fit the parameters of the mass profile ($r_{200},\,r_\text{s}$), the anisotropy parameter ($\beta_{\infty}$) and the modified gravity parameter $Y_1$. 

%\footnote{Note that we choose to compute confidence intervals using $\chi^2$ levels instead of directly integrating the likelihood. This is because $\chi^2$ regions are not sensitive to the peak of the distribution, but to the overall shape of the likelihood, which in our case may exhibit irregular features. Moreover, for non-normalizable   distributions, $\chi^2$ bounds can be always set, independently of the integration limits, which is advantageous for the marginalisation process, although in this case one cannot associate a probability meaning to the likelihood.}.
For the velocity anisotropy, from now on we use
\begin{equation}
 \mathcal{A_\infty} \equiv (1-\beta_{\infty})^{-1/2}=\left(\frac{\sigma_r}{\sigma_\theta}\right)_{r\to\infty},   
\end{equation}
as in e.g. \citet{Mamon10}, which is equal to unity for isotropic distributions ($\beta_{\infty}=0$) and to $1.41$ for $\beta_\infty=0.5$.
As explained in Section \ref{sec:Codes}, the fitting procedure can be performed by computing the likelihood over a multi-dimensional grid or by means of an MCMC algorithm. In this work, in order to explore more efficiently our range of parameters, we carry out the MCMC sampling over $3\times 10^5$ points in the space ($\log(r_{200}),\log(r_\text{s}),\log(\mathcal{A}_\infty),$ $Y_1$). Note that we have considered the logarithm of the parameters, except for $Y_1$ which can be negative, in order to give an equal weight to all values. We assume flat priors in the allowed ranges of values for each parameter, listed in Table \ref{tab:ranges}. The lower bounds of $Y_1>-2/3\simeq -0.67$ follows from the stability condition discussed in \citet{Babichev:2016jom}  

\begin{figure*}
\centering
\includegraphics[scale=0.5]{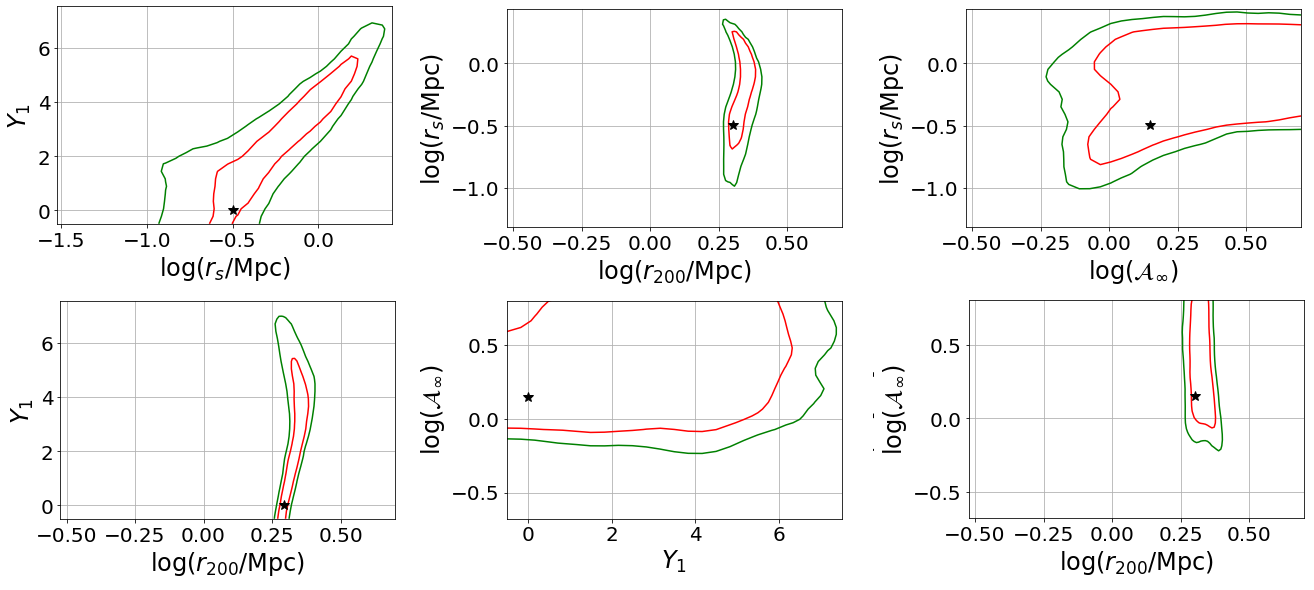}
\caption{\label{fig:horn1}Iso-chi square contours of the marginalised 2-dimensional distributions for the free parameters in the \textsc{MG-MAMPOSSt} fit, i.e the NFW profile and modified gravity parameter $Y_1$. The results are shown for one halo in the analysis, but they are representative of the behaviour of all the cluster sample. Green curves: $2\Delta(\ln\mathcal{L})=6.1$ contours. Red curves: $2\Delta(\ln\mathcal{L})=2.3$. The black stars indicate the true values of the halo's parameters.}
\end{figure*}

\begin{figure}
\centering
\includegraphics[width=\columnwidth]{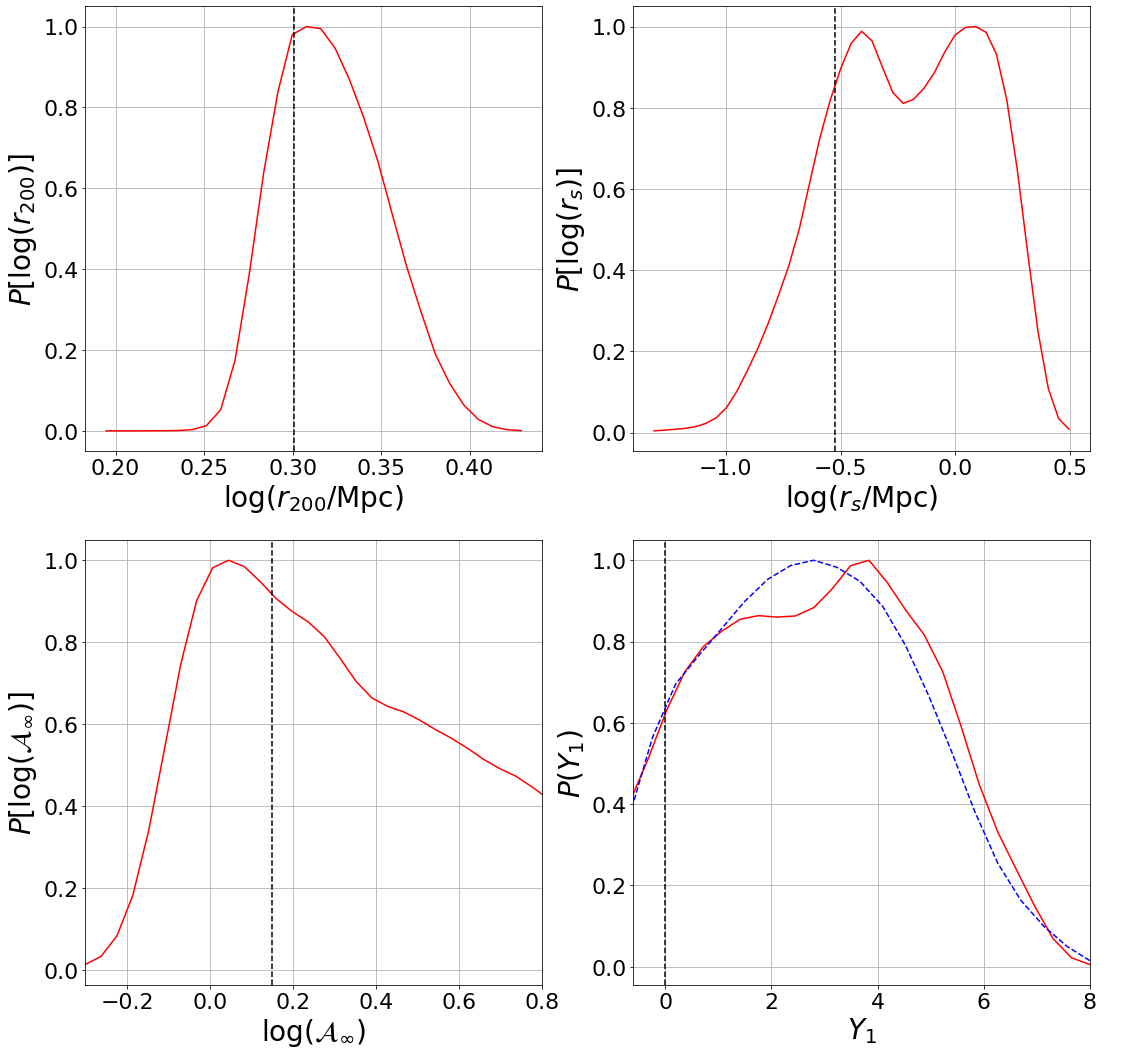}
\caption{\label{fig:horn2} Marginalised one-dimensional distributions of the four free parameters obtained by the \textsc{MG-MAMPOSSt} analysis for the same halo of Fig. \ref{fig:horn1}. The vertical dashed lines represent the true value of the parameters while the blue dashed curve in the bottom right plot shows the marginalized distribution of $Y_1$ obtained when considering a halo with $r_{200}=1.4\,\text{Mpc}$ and $r_\text{s}=0.33\,\text{Mpc}$.}

\end{figure}

The results for the 2-dimensional marginalised iso-probability contours $2\Delta(\ln\mathcal{L})$, where $\mathcal{L} \equiv \mathcal{L}^\text{dyn}$ is the \textsc{MG-MAMPOSSt} likelihood, and the one-dimensional distributions for each parameter are shown in Fig. \ref{fig:horn1} and Fig. \ref{fig:horn2} respectively, for a single cluster.

A first important lesson is that the coupling constant $Y_1$ is strongly degenerate with the mass profile parameters -- increasing the NFW parameters $r_\text{s}$ and $r_{200}$ corresponds to an increase of $Y_1$ respectively. This is expected, since increasing $Y_1$ leads to a lower gravitational potential gradient (compared to GR) for $r>r_\text{s}$, which can be compensated by assuming either a larger $r_{200}$ or a larger $r_\text{s}$. As a consequence, the effect of a standard Newtonian gravitational potential in a galaxy cluster phase space can be mimicked to a certain extent by introducing an additional degree of freedom $Y_1$ and suitably adjusting the values of the other parameters. This aspect needs to be accounted for in the analysis of real observations when placing constraints on the modified gravity coupling.

Fig. \ref{fig:horn2} shows the marginalised, 1-dimensional likelihoods for one halo in the sample. Note that all the distributions of $(Y_1,\,r_\text{s},\,r_{200})$ are shifted towards larger positive values with respect to the expectation, particularly for $Y_1$. 
This is a statistical artifact and not a physical effect -- indeed, all the four-dimensional likelihoods exhibit a maximum in the neighborhood of the halo's parameters true values (i.e. $Y_1=0$, $r_{200}=2.0\,\mpc$, $r_\text{s}=0.3\,\mpc$, $\mathcal{A}_\infty=1.41$) by construction. However, they also have a second local maximum at $Y_1>1$, due to the strong degeneracy with the mass profile parameters, which shifts the 1-dimensional distribution  when marginalizing. We can always set an upper limit  $Y_1 < 6.92$ at $2\Delta(\ln\mathcal{L})=4.0$ from the marginalised distribution (corresponding to 95\% C.L.). The value has been chosen as the largest constraint obtained among the 20 marginalised distributions $P(Y_1)$ of each of the synthetic phase spaces in the sample. 
In the bottom right plot of Fig. \ref{fig:horn2} we have also shown the marginalized distribution obtained from the \textsc{MG-MAMPOSSt} analysis of a halo with $r_{200}=1.4\,\text{Mpc}$ and $r_\text{s}=0.33\,\text{Mpc}$ (i.e. smaller concentration and smaller mass with respect to the reference case). As expected, the two distributions are basically equivalent, modulo statistical fluctuations. This confirms that this modified gravity model is not affected by the size (or mass) of the cluster.

\subsubsection{Joint clusters}
\begin{figure}
\centering
\includegraphics[width=.8\columnwidth]{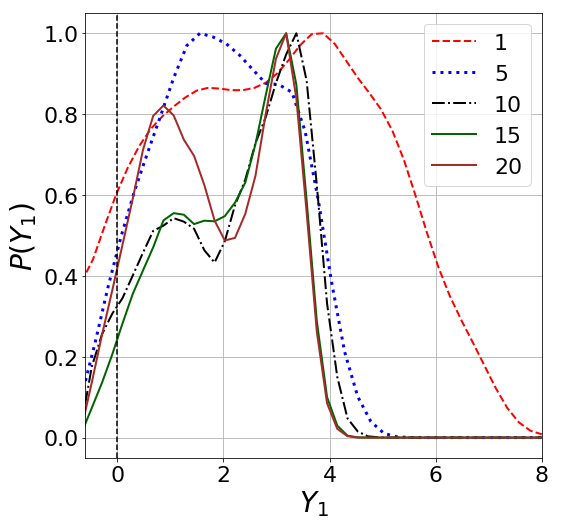}
\caption{\label{fig:horncomb5} Marginalised distributions $P(Y_1)$ obtained by combining likelihoods of single haloes according to Eq. \eqref{eq:liketot}. Different curves correspond to different number of phase spaces considered in the combination. The effect of strong degeneracy between $Y_1$ and the NFW parameters $r_\text{s},\,r_{200}$, is particularly evident for $N_\text{haloes}\gtrsim 10$, giving rise to a spurious peak at $Y_1>0$.} %The value of $P(Y_1)$ corresponding to the second maximum, near the GR expectation ($Y_1 = 0$), increases with the number of haloes but is always subdominant with respect to the mode of the distribution.}

\end{figure}

So far we focused on a single cluster. Still working with the synthetic clusters constructed assuming GR, we combine the individual likelihoods of each phase space  to study the joint distribution as a function of the number of haloes. We write,
\begin{equation}\label{eq:liketot}
    -\ln\mathcal{L}_{\text{tot}}=\sum_i^{N_\text{h}}\left(-\ln\mathcal{L}^{\text{dyn}}_i\right),
\end{equation}
where $N_\text{h}$ is the number of phase spaces in the sum. Remember that we are considering realizations of haloes characterized by the same NFW parameters (i.e. the values of $r_{200},\,r_\text{s}$ and $\mathcal{A}_\infty$ are the same for each realization), but with distinct projected phase spaces. Thus, a stacked analysis can be done by Monte-Carlo sampling over the combination of single likelihoods  of different (synthetic) clusters as in Eq. \eqref{eq:liketot}. 
In Fig. \ref{fig:horncomb5} we plot five marginalised distributions $P(Y_1)$, corresponding to different number of haloes $N_\text{h}$ taken into account in Eq. \eqref{eq:liketot}. 
The curves exhibit an apparent tension with $Y_1=0$ for $N_\text{h}>1$. 
When increasing  the number of haloes, the distributions develop a second maximum near the GR expectation value which, however, remains subdominant with respect to the first peak. As one may naively expect, this confirms that the statistical degeneracy, responsible of the spurious maximum in the marginalised likelihood, cannot be removed just by increasing the number of haloes in the fit. The constraints obtained at $95\%$ C.L. are listed in the second column of Table \ref{tab:Y}.
For $N_\text{h}\sim 20$ we may extract, at $95\%$ C.L., an upper limit $Y_1 \lesssim 3.9$ and a lower limit $Y_1 \gtrsim - 0.44$, marginally close to the theoretical bound, $Y_1 > - 2/3$ (e.g. \citealt{Saltas:2018mxc,Sakstein:2018fwz,Creminelli:2018xsv}).

The results of our analysis indicate that, in the case of VS models, the constraints obtainable with the \textsc{MG-MAMPOSSt} fit alone are limited by the statistical degeneracy between model parameters. Other probes, such as  X-ray and lensing measurements, are required in order to remove the unphysical region of the parameter space associated to large values of $r_\text{s},\,r_{200},\,Y_1$. Similar conclusions have been reached by Paper I for the application of \textsc{MG-MAMPOSSt} to linear $f(\mathcal{R})$ gravity.
Therefore, although kinematics alone are not able to provide sufficiently constraining bounds on $Y_1$, one may expect to obtain valuable results from the combination of several data-sets obtained with different methods. In order to further investigate this point, in the following we simulate the availability of additional lensing information to understand at what level of precision the  models can be constrained with our method. 

\begin{table} 

\begin{tabular}{|c|c|c||}
\hline

Parameter & Min value allowed & Max value allowed \\
\hline
$\log(r_{200})$ & $ \log(0.15\,r^\text{true}_{200})$ & $\log(2.5\,r^\text{true}_{200})$    \\ 

$\log(r_\text{s})$ & $ \log(0.10\,r^\text{true}_\text{s})$ & $ \log(8.0\,r_\text{s}^\text{true})$ \\ 

$\log(\mathcal{A}_\infty)$ & $ \log(0.15\,\mathcal{A}^\text{true}_\infty)$ & $ \log(5.0\,\mathcal{A}^\text{true}_\infty)$\\  
 $Y_1$    & $-0.67$ & $8.0$  \\ 
 $Y_2$   & $-8.0$ & $8.0$  \\ 
$\mathcal{Q}_2$ & $0$ & $1$  \\ 
$\phi_2$ & $ 0 $ & $ 1 $ \\ 
\hline
\end{tabular}
\caption[Parameter range]{\label{tab:ranges} Allowed parameter range in the \textsc{MG-MAMPOSSt} fit used for all the analyses presented in this paper. The quantities $r^\text{true}_{200},\,r^\text{true}_\text{s}$ and $\mathcal{A}^\text{true}_\infty$ indicate the true values of the cluster's parameters. $\mathcal{Q}_2$ and $\phi_2$ are defined according to Eq. \eqref{eq:scaledvar}. In the case of VS, all the combinations of  $Y_1$ and $Y_2$ within the ranges giving rise to a negative effective mass have been excluded form the analyses.}

\end{table}

\subsection{Joint Lensing and internal kinematics analysis}

%\textcolor{red}{This is just a try for this very ugly function:}
%\begin{widetext}
%\begin{equation}
%\begin{array}{cc}
% f_\text{MG}(x)= 
%\begin{cases}
%  \,\,\,\mathcal{P}_1(x,Y_1,Y_2)\left(x^2-1\right)^{-3} + \text{sech}^{-1}(x)\, \mathcal{P}_2(x,Y_1,Y_2)\left(1-x^2\right)^{-7/2} & x<1 \\
%  \\
%  \,\,\,\mathcal{P}_1(x,Y_1,Y_2)\left(x^2-1\right)^{-3}-\sec ^{-1}(x) \mathcal{P}_2(x,Y_1,Y_2)\left(x^2-1\right)^{-7/2} & x>1 \\
%  \\
% \,\,\,(35-10\, Y_1-70\, Y_2)/105 & x=1 \\
%\end{cases} 
%\end{array}
%\end{equation}
%\end{widetext}
%Where: 
%\begin{equation*}
%   \mathcal{P}_1(x,Y_1,Y_2)= \frac{1}{2}\left[8 x^4+x^2 (10 Y_1+15 Y_2-16)+5 Y_1-15 Y_2+8\right]
%\end{equation*}
%\begin{equation*}
% \mathcal{P}_2(x,Y_1,Y_2)=\frac{1}{2}\left[x^4 (2 Y_1+5 Y_2+8)+\right.
%\end{equation*}
%\begin{equation}
%    \left.x^2(11 Y_1+5 Y_2-16)+2 (Y_1-5 Y_2+4)\right]
%\end{equation}

Present and upcoming high precision strong+weak lensing measurements can recover the mass profile parameters $r_{200}$ and $r_\text{s}$ with uncertainties down to $\sim 7-10\%$ and $\sim 30\%$ respectively (e.g. \citealt{Umetsu16}). In order to forecast the constraints obtainable by our method with an additional lensing probe, we need to correctly simulate a full convergence and shear profile for our mock haloes as well as to employ a strong lensing modeling for the inner core of the cluster. Instead of implementing a full ray-tracing simulation, here we focus only on weak lensing analyses by building a mock (reduced) tangential shear profile keeping our ideal assumption of perfect control of systematic effects.

As mentioned before, we aim at applying the work presented in this paper in order to obtain constraints from a joint lensing and kinematic analysis of the relaxed massive cluster MACS 1206, for which accurate strong+weak lensing profiles has been derived in GR (\citealt{UmetsuMACS,Umetsu16,Caminha2017}). As such, we calibrate our lensing model over an NFW halo characterised by $r_{200}=2.0\,\mpc$ and $r_\text{s}=0.3\,\mpc$, located at the same redshift of MACS 1206, $z_c=0.44$.

In the weak-lensing approximation, the distortion of the background sources is relatively small (see e.g. \citealt{Kaiser95}) and the reduced tangential shear profile can be defined as:
\begin{equation}
    g_\text{t}(R)=\frac{\gamma_\text{t}(R)}{1-\kappa(R)},
\end{equation}
where $\kappa$ and $\gamma_\text{t}$ represent the convergence and tangential shear profiles as a function of the projected radius $R$. These quantities can be expressed in terms of the lens' surface mass density 
\begin{displaymath}
\kappa(R)=\frac{4\pi G}{c^2_{\rm l}}\frac{D_\text{l}D_\text{ls}}{D_\text{s}}\Sigma(R),
\end{displaymath}
\begin{equation}
 \gamma_\text{t}(R)=\frac{4\pi G}{c^2_{\rm l}}\frac{D_\text{l}D_\text{ls}}{D_\text{s}}\left[\bar{\Sigma}(R)-\Sigma(R)\right], 
\end{equation}
where $D_\text{s}(z_s),\, D_\text{l}(z_c)$ are the angular diameter distances of a source located at redshift $z_s$ and of the lens at the cluster redshift $z_c$, respectively. $D_\text{ls}(z_c,z_s)$ is the angular diameter distance between the source and the lens.
Finally, $\bar{\Sigma}(R)=M_\text{proj}(R)/(\pi R^2)$ is the average surface mass density profile. The surface mass density is obtained with the projection equation: 
\begin{equation} \label{eq:sigmaNFW}
    \Sigma(R)=2\int_R^\infty{\rho(r)\frac{r}{\sqrt{r^2-R^2}}\text{d}r},
\end{equation}
while the projected mass, $M_\text{proj}(R) =\int_0^R 2\pi S \,\Sigma(S)\,\text{d}S$
involves a double integral, which can be expressed with single integrals as
\begin{equation} \label{eq:MpjNFW}
    M_\text{proj}(R)=4\pi\left[\int_0^R{\rho(r)r^2\text{d}r}+\int_R^\infty{\rho(r)\left( r^2-r\sqrt{r^2-R^2}\right)\text{d}r}\right],
\end{equation}
(Appendix A of \citealt{Mamon10}).
Analytical expressions for $\Sigma(R)$ and $M_\text{proj}(R)$ for the NFW
    model were first derived by \citet{Bartelmann96}. 

In order to build our mock profile, we should assume a distribution of the background sources to average the lensing convergence and tangential shear over all source redshifts. Following e.g. \citet{Chen20} and references therein, we consider a density of sources described by
\begin{equation}
n(z_s)=\mathcal{C}\,z_s^2\exp\left(-\frac{z_s}{z_0}\right),\,\,\,\,\,\,\,\, z_0=\frac{1}{3}\left(\frac{n_\text{g}}{30}\right)^{1/3}.
\end{equation}
In the above equation, $\mathcal{C}$ is a normalization factor and $n_\text{g}$ is the average number of source galaxies per $\text{arcmin}^2$ observable from current and future surveys. Here we assume $n_\text{g}=30 \, \text{arcmin}^{-2}$, as expected for the Wide Survey of the Euclid mission (\citealt{laureijs11}). Furthermore, we found that varying $n_\text{g}$ in a reasonable range -- from $20\, \text{arcmin}^{-2}$ to $40\, \text{arcmin}^{-2}$ -- produces negligible effects on our results. For $n_\text{g}=30$, the density distribution peaks at $z_s\sim 0.6$ and sharply decreases at higher redshift becoming negligible for $z\gtrsim 3$.
The reduced tangential shear profile, averaged over all redshifts, is then given by
\begin{equation}
    \langle g_\text{t} \rangle \simeq \frac {\langle \gamma_\text{t} \rangle}{1- f_\text{l}\langle \kappa \rangle},
\end{equation}
where the average $\langle X \rangle$ is obtained by integrating $X$ weighted with $n(z_s)$ from $z_c$ to the maximum redshift of the sources $z_m$. In the following, we set $z_m=10$. 
The factor $f_\text{l}$, of order unity, is given by the ratio $\langle W^2 \rangle/\langle W \rangle^2$, with $W=D_\text{ls}/D_\text{s}$.

We consider $N_\text{b}=10$ log-spaced radial bins from the cluster center over the angular range $[0.6', 16']$. For the lower bound, we choose a value which is larger than the Einstein radius of the cluster ($\sim 28''$ for MACS 1206, see e.g. \citealt{UmetsuMACS,Zitrin2015}) to ensure the validity of weak lensing approximation. The upper limit is set following the weak lensing analysis of 16 X-ray selected clusters by \citet{Umetsu14}. To each radial bin we assign an associated error $\sigma_{\text{l},i}$ given by the quadratic sum of two contributions
\begin{equation}\label{eq:errorlens}
\sigma^2_{\text{l},i}=\sigma^2_{\text{e},i}+\sigma^2_\text{lss},
\end{equation}
where $\sigma^2_{\text{e},i}=\sigma^2_\text{g}/[\pi(\alpha^2_\text{up}-\alpha^2_\text{low})n_\text{g}]$ is the noise due to the intrinsic ellipticity $\sigma^2_\text{g}$ of the sources lying within an annulus between the angles $\alpha_\text{low}$ and $\alpha_\text{up}$, and $\sigma^2_\text{lss}$ expresses the effect of the uncorrelated projected large scale structure \citep{Hoekstra03}. The full covariance matrix due to the cosmic-noise can be estimated numerically as shown by e.g. \citet{umetsu11} and the computation requires the knowledge of the non-linear matter power spectrum. The overall result is of the order of a few thousandths and the contribution of $\sigma^2_\text{lss}$ becomes relevant only at large cluster radii (see Figure 14 of \citealt{Umetsu_2020}). Working in a simplified picture, we assume a constant value $\sigma_\text{lss}=0.005$ for all bins. As for $\sigma_\text{g}$, typical values of  adopted in the literature vary between $\sim 0.25$ and $\sim 0.4$ (see e.g. \citealt{Kohlinger15,Chen20}). In our simulation we assume $\sigma_\text{g}=0.3$.

\begin{figure}
\centering
\includegraphics[width=\columnwidth]{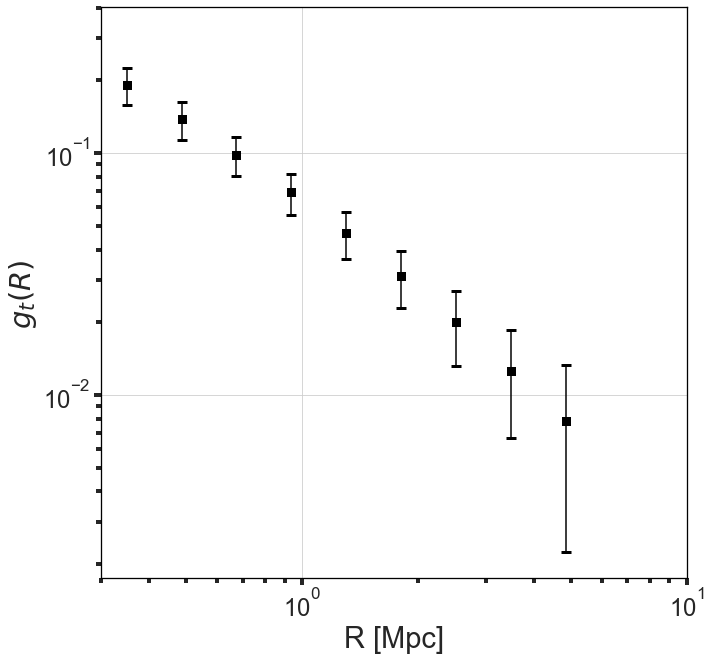}
\caption{\label{fig:tan_shear} Mock tangential shear profile built for the reference NFW halo in our analysis, $r_\text{s}=0.3\,\mpc$, $r_{200}=2.0\,\mpc$. The error bars indicate the values of $\sigma_{\text{l},i}$. The relative increase of the error bars at larger radii is due to the dominance of uncertainties associated to large-scale structures, as explained in the text.}

\end{figure}

The binned mock tangential shear profile for our reference NFW halo is shown in Figure \ref{fig:tan_shear} as a function of the projected radius $R=\alpha D_\text{l}(z_c)$.

In Section \ref{sec:theo} we have mentioned that the lensing Poisson equation relates the effective mass density profile to the sum of the gravitational potentials.
In VS models, the effective density $\rho_\text{lens}$ is given by Eq. \eqref{eq:BH_efflens}. Assuming a NFW model, this becomes:
\begin{equation}
    \rho_\text{lens}(r)=\rho_\text{NFW}(r)\left[1+\frac{Y_1-5Y_2-(2Y_1+5Y_2)\,r/r_\text{s}}{4\,(1+r/r_\text{s})^2}\right],
\end{equation}
which can be further integrated by means of Eqs. \eqref{eq:sigmaNFW}, \eqref{eq:MpjNFW} in order to obtain the effective tangential shear profile $g_{\text{t,vs}}(R_i|\bm{\theta}_\text{l})$ as a function of the parameter vector $\bm{\theta}_\text{l}=(r_{200},r_\text{s},Y_1,Y_2)$.
For each choice of the free parameters, the lensing (log) likelihood is then computed as:
\begin{equation}
    \ln\mathcal{L}_\text{lens}(\bm{\theta}_\text{l})=-\frac{1}{2}\sum_{i=1}^{N_\text{b}}\frac{\left[\langle g_\text{t}(R_i)\rangle-\langle g_{\text{t,vs}}(R_i|\bm{\theta}_\text{l})\rangle\right]^2}{\sigma^2_{\text{l},i}},
\end{equation}
where $N_\text{b} = 10$ is the number of bins.

As done above, we perform a MCMC sampling in the space ($\log(r_{200}),\log(r_\text{s}),\log(\mathcal{A}_\infty),$ $Y_1,Y_2$), assuming a flat prior for each parameter within the allowed ranges, which are listed in Table \ref{tab:ranges}. When $Y_1$ and $Y_2$ are considered, we excluded all the combinations for which the resulting mass profiles become negative.
Moreover, $Y_1$ is assumed to be strictly larger than $-2/3$ to fulfill the stability conditions highlighted earlier. The sampling has been performed by using the joint distribution
\begin{displaymath}
\ln\mathcal{L}_{\text{joint}}=\ln\mathcal{L}_{\text{tot}}(r_\text{s},r_{200},\mathcal{A}_\infty,Y_1)+N_\text{h}\ln\mathcal{L}_\text{lens}(\bm{\theta}_\text{l}),
\end{displaymath}
where $N_\text{h}$ indicates the number of clusters considered in the stacking analysis and  $\mathcal{L}_{\text{tot}}(r_\text{s},r_{200},\mathcal{A}_\infty,Y_1)$ is the $N_\text{h}$ clusters-combined \textsc{MG-MAMPOSSt} likelihood, Eq. \eqref{eq:liketot}. 
 
\begin{figure}
\centering
\includegraphics[width=\columnwidth]{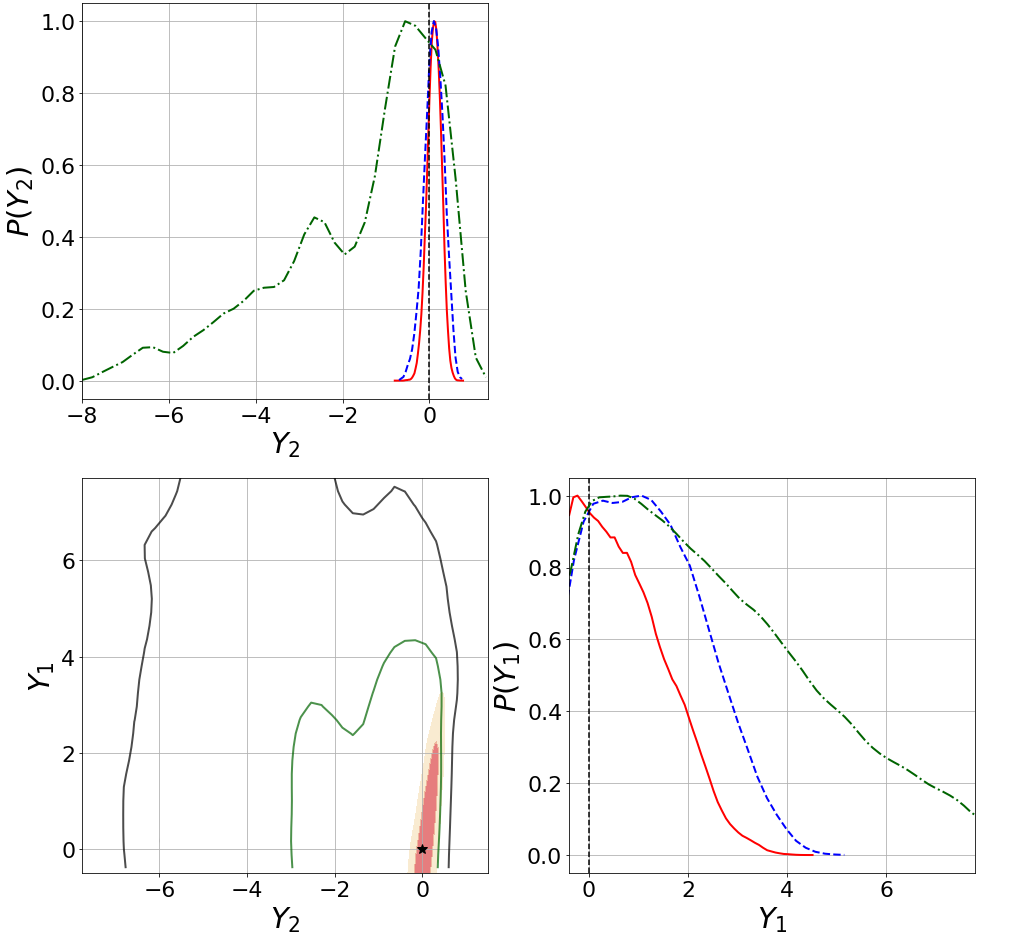}
\caption{\label{fig:lensing1} Bottom-left plot: two-dimensional distribution in the space $(Y_1,Y_2)$ obtained from the lensing+kinematics likelihood of one cluster with $N=600$ tracers in the \textsc{MG-MAMPOSSt} fit. The dark red and light red shaded areas represent the $1\sigma$ and $2\sigma$ contours, respectively, while the green and black contours refer to the same 
for lensing analysis only. Upper and bottom-right plots: marginal distributions of $Y_1$ and $Y_2$ for the joint lensing+kinematics analysis of the same cluster (red lines), for a cluster with $N=100$ tracers (blue dashed curves), and for lensing only (green dash-dotted curves).} 
The black vertical lines indicate the GR expectation $(Y_1=Y_2=0)$. 

\end{figure}

The constraints  at 95\% C.L. on the VS parameters $Y_1$ and and $Y_2$ are listed in Table \ref{tab:Y} for different number of clusters considered in the combined analysis. In Fig. \ref{fig:lensing1} we show the marginalized posterior distributions of $Y_1$ and $Y_2$ for a single halo in the sample when $\sim 600$ (red solid lines) and $\sim 100$ (blue dashed lines) tracers are considered in the \textsc{MG-MAMPOSSt} fit, compared with the posteriors from lensing only (green dash-dotted curves). Adding the internal kinematics information to the lensing posterior dramatically
improves the constraint on $Y_2$ and strongly improves that on $Y_1$. For the joint kinematic+lensing analysis of the reference case ($N=600$), we obtain at 95\% C.L.:
\begin{equation}
    Y_1\lesssim 2.75, \,\,\,\,\,\,\,\,\,\, Y_2=-0.08^{+0.32}_{-0.28}.
\end{equation}
Note that for $Y_1$ we consider only the upper limit, as the lower bound is still given by the value settled by the stability conditions  $Y>-2/3$.

\begin{table*} 
\centering
\begin{tabular}{|c|c|c|c|c|c|c|c|c|c|}
\hline
&  \multicolumn{7}{c|}{Vainshtein screening} &\multicolumn{2}{c|}{$f(\mathcal{R})$ gravity} \\
\cline{1-5}
\cline{6-10}
&  {$N=600$ (MAM)}  & \multicolumn{2}{c|}{$N=600$ (lens)} & \multicolumn{2}{c|}{$N=600$ (joint)} & \multicolumn{2}{c|}{$N=100$   (joint)} &  $N=600$  & $N=100$  \\ 
\cline{1-10}

$N_\text{h}$ clusters & $Y_1$ & $Y_1$ & $Y_2$ & $Y_1$ & $Y_2$ & $Y_1$ & $Y_2$ & $|f_{\mathcal{R}0}|$ & $|f_{\mathcal{R}0}|$ \\
\hline
& & & & &  & &  & & \\

1 & $\lesssim 6.92$ & $\lesssim 7.45$ & $-0.72^{+1.55}_{-4.18}$ & $ \lesssim 2.75$ & $0.08^{+0.32}_{-0.28} $ &  $ \lesssim 3.56$ &  $0.10^{+0.44}_{-0.40} $ & -- & -- \\
& & & & &  & &  & & \\
5 & $ \lesssim 4.10$ & $\lesssim 3.62$ & $0.04^{+0.52}_{-1.05}$ & $ \lesssim 1.65$ & $0.06^{+0.20}_{-0.18}$ &  $\lesssim 1.87$ & $-0.08^{+0.31}_{-0.20}$ & $\lesssim 3.37\times 10^{-5}$ & $\lesssim 5.13\times 10^{-5}$\\ %5.05
& & & & &  & &  & & \\
10 & $ 3.52^{+0.62}_{-3.99}$ & $\lesssim 2.90$ & $0.04^{0.42}_{0.63}$ &$ \lesssim 1.24$ & $-0.05^{+0.17}_{-0.13}$ & $\lesssim 1.65$ & $0.01^{+0.24}_{-0.17}$ & $\lesssim 1.12\times 10^{-5}$ & $\lesssim 3.24\times 10^{-5}$\\
& & & & &  & &  & & \\
15 & $ 3.18^{+0.70}_{-3.64}$ & $\lesssim 2.44$ & $0.07^{+0.49}_{-0.38}$ &  $ 0.04^{+1.00}_{-0.39}$ & $0.01^{+0.12}_{-0.09}$ & $\lesssim 1.20$ & $-0.01^{+0.19}_{-0.16}$ & $\lesssim 9.51\times 10^{-6}$ & $\lesssim 2.43\times 10^{-5}$\\ %4.50
& & & & &  & &  & & \\
20 & $ 3.17^{+0.69}_{-3.61}$ & $\lesssim 2.22$ & $0.07\pm 0.38$ &  $0.08^{+0.77}_{-0.34}$ & $0.01^{+0.09}_{-0.08}$ & $\lesssim 1.02$ & $0.01^{+0.16}_{-0.14}$ & $\lesssim 7.11\times 10^{-6}$ & $\lesssim 1.79\times 10^{-5}$\\
\hline

\end{tabular}
\caption[Constraints on $Y_1$ and $Y_2$]{\label{tab:Y} Constraints at 95\% C.L forecasted from the \textsc{MG-MAMPOSSt} method with additional lensing information for the parameters of the two modified gravity models presented in this paper, in the case of clusters with $r_{200}=2.0\,\mpc$ and $r_\text{s}=0.3\,\mpc$ ($c\simeq 6.67$). From column two to column eight: Vainshtein screening. Column nine and ten: chameleon $f(\mathcal{R})$ gravity. For both models we show the results when using $\sim 600$ and $\sim 100$ cluster members in the \textsc{MG-MAMPOSSt} fit.}
\end{table*}

As expected, our results are consistent with the GR predictions and of the same order of current cosmological bounds on those parameters derived with similar methods.
In particular,   \citet{Sakstein:2016ggl} obtained constraints of order unity on $Y_1$ and $Y_2$ by combining the stacked X-ray surface brightness profiles from the XMM Cluster Surveyand weak lensing profiles from CFHTLenS of 58  clusters at redshift $0.1 < z < 1.2$. A much more stringent constraint has been found by  \citet{Salzano17} who however worked on a quite different class of models where the potential $\Phi$ and $\Psi$ are modified in terms of one single free parameter $Y_1\equiv Y_2=\mathcal{Y}$ (although is still true that $\Phi \ne \Psi$). By combining X-ray mass profiles and strong+weak lensing analyses for 18 clusters studied within the CLASH collaboration, they found a very tight upper limit of $\mathcal{Y}\lesssim 0.16$ at $2\,\sigma$. 

For the general case $Y_1\ne Y_2$, our simplified approach shows that the parameter $Y_2$ can be constrained at a level of $30-40\%$ with a single cluster, down to $\sim 10\%$ for the combination of 20 clusters. However, the constraints on $Y_1$ are still one order of magnitude larger than what is found by other astrophysical probes. By focusing on the same class of theories as in \citet{Salzano17}, we forecast $\mathcal{Y}\lesssim 0.26$ at 95\% C.L. from the analysis of a single halo.
In view of the above results, we remark the following: 
\begin{itemize}
    \item Our forecasts are obtained by assuming an ideal situation with perfect control of systematics for both lensing and internal kinematics analyses.  
    \item The uncertainties considered here are reliable, given the current available data-sets, but slightly optimistic. Indeed, high-precision spectroscopic measurements for $\sim 600$ cluster members could be obtained only for a small fraction of observed galaxy clusters. Therefore, we consider a more realistic situation of $\sim 100$ tracers in the \textsc{MG-MAMPOSSt} fit. The results are listed in the seventh and eighth column of Table \ref{tab:Y}. As expected, in this case the constraints are $\sim 25\%$ weaker, but still good for $Y_2$ even for a single halo, confirming that the availability of lensing information plays a crucial role in the \textsc{MG-MAMPOSSt} analysis.

 \item The test carried out shows that galaxy cluster mass profiles are not able to provide tight constraints for the Newtonian potential $\Phi$ for VS (related to the coupling $Y_1$), compared to current astrophysical constraints, although the method offers a complementary test for these models at cosmological scales. Nevertheless, the combination of dynamics+lensing analyses in clusters promises to provide very useful information on the parameter $Y_2$ in VS models, for which the only constraint obtained at cosmological level is that of \citet{Sakstein:2016ggl}. Indeed, the effective lensing mass, Eq. \eqref{eq:lensmass}, for a NFW density profile reads as,

\begin{equation*} 
    M_{\text{lens}}=M_{\text{NFW}}+\frac{r^2M_{200}\left[Y_1(r_\text{s}-r)-5Y_2(r_\text{s}+r)\right]}{8[\ln(1+c)-c/(1+c)]}\frac{1}{(r_\text{s}+r)^{3}}
\end{equation*}

\begin{equation}\label{eq:Mlens}
    \equiv M_{\text{dyn}}+M_2, 
\end{equation}
with 
\begin{equation}
   M_2=\frac{r^2M_{200}}{8(r_\text{s}+r)^{3}}\frac{Y_1(r-r_\text{s})-5Y_2(r_\text{s}+r)}{[\ln(1+c)-c/(1+c)]}. 
\end{equation}
We can see that the contribution of $Y_2$ to departures from GR is five time stronger than $Y_1$. Moreover, the last term in Eq. \eqref{eq:Mlens} is always different from zero for any finite $r$, as shown by the relative ratios of the lensing mass profile in Fig. \ref{fig:relY2}.  Thus, we expect that joint lensing and dynamics analyses can constrain much better $Y_2$ than $Y_1$. 
\begin{figure}
\centering
\includegraphics[scale=0.45]{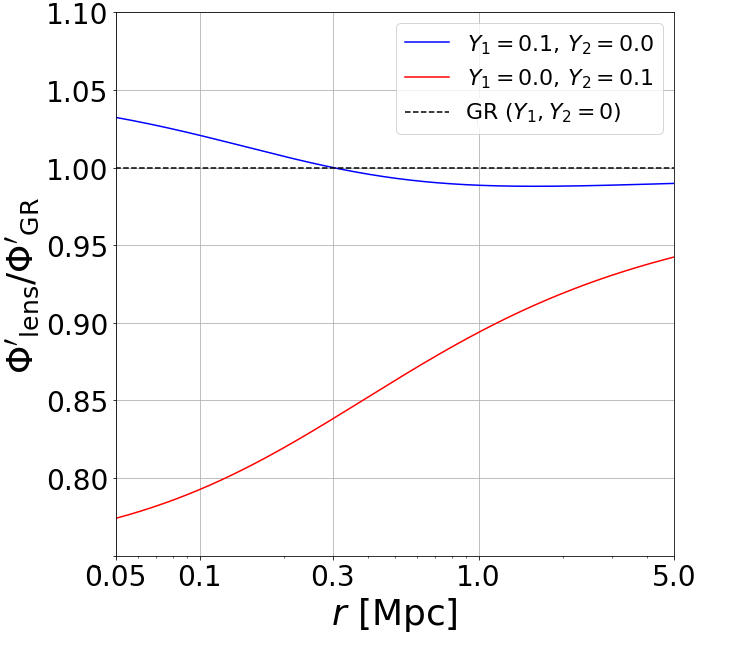}
\caption{\label{fig:relY2} Ratios between the gradient of the lensing potential $\Phi'_\text{lens}$ and the gradient of GR gravitational potential $\Phi'_\text{GR}$, obtained for a VS model with $Y_1=0.1,\,Y_2=0$ (blue) and $Y_1=0,\,Y_2=0.1$ (red). The NFW profile parameters are $r_{200}=2.0\,\text{Mpc}$ and $r_\text{s}=0.3\,\text{Mpc}$.}
\end{figure}

In this perspective, we further outline that:
    
    \item The method we used to reconstruct the parameters from the simulated lensing mass profile only accounts for weak lensing (mock) data. In general, high precision joint strong+weak lensing analyses - like those performed by  \citet{Umetsu16,Caminha2017} - can determine the mass profile down to the cluster core. The structure of Eq. \eqref{BH-NFW} and Eq. \eqref{eq:Mlens} suggests that the knowledge of the inner shape of the profile provides significant  information in breaking the degeneracy between the model parameters.
    
\end{itemize}

\section{Application II: chameleon screening}\label{sec:fr}
In this section, we present the \textsc{MG-MAMPOSSt} results for the sample of mock haloes assuming a fifth force mediated by a
chameleon scalar field $\phi$.

\subsection{General chameleon screening}
In the general case assuming that the interior solution for the scalar field is close to zero up to the screening radius, the magnitude of the additional force is totally determined by the background field value $\phi_{\infty}$ and the coupling constant $\mathcal{Q}$.
We first analyse our sample of synthetic phase spaces in this framework, using Eq. \eqref{eq:field} to model the chameleon profile and assuming $\phi_\text{int}\simeq 0$. We perform a MCMC sampling of the likelihood from the internal cluster kinematics $\mathcal{L}_{\text{dyn}}$ over the full parameter space $(\log r_\text{s},\, \log r_{200},\, \log \mathcal{A}_\infty,\phi_\infty,\mathcal{Q})$. 
For each parameter we assume flat priors in the allowed range, whose upper and lower bounds are listed in Table \ref{tab:ranges}. Note that for the GR kinematic parameters $(r_\text{s},\, r_{200},\, \mathcal{A}_\infty)$ we used the same values as in the VS case.

In principle, while the CS mechanism recovers GR at small scales, modifications of gravity are further Yukawa-suppressed for $r\gtrsim \lambda_\phi\equiv m^{-1}_\phi \sim \left[\mathcal{Q} \mathcal{R}_0 M_\text{P}c_\text{L}^2/\phi_\infty\right]^{-1/2}$, where $\lambda_\phi$ is the background interaction range of the force, $\mathcal{R}_0$ is the background Ricci scalar.
In order to satisfy local test of gravity at Solar System scales, it has been shown that $\lambda_\phi$ should be of the order of few $\text{Mpc}$ (see e.g. \citealt{Wang13}). 
In  this simple treatment we neglect the Yukawa suppression of the chameleon field as in the \textsc{MG-MAMPOSSt} analysis we consider only galaxies lying within a projected radius $R\sim r_{200} \sim \mathcal{O}(\text{Mpc})$. However, since in the Jeans equation the integral of the velocity dispersion extends far beyond the viral radius, for large $\mathcal{Q}$ the motion of member galaxies could be affected by the contribution of the tails of the integral even if the cluster is totally screened (i.e. $S\simeq r_{200}$). In order to avoid this problem, we make a conservative assumption that the effect of the chameleon field is zero everywhere if $S >1.5\,r_{200}$. A complete solution of the field equation will be investigated elsewhere.

Since in CS the lensing mass profile is unaffected by the fifth force, the contribution of additional lensing information affects only the NFW mass profile parameters $r_{200}$, $r_\text{s}$. Thus, in this case we follow the simpler approach of \citet{Pizzuti:2019wte} who considered a Gaussian distribution $P_\text{lens}(r_\text{s},r_{200})$ combined with the \textsc{MG-MAMPOSSt} likelihood. The additional distribution is assumed to be derived by an unbiased (lensing) reconstruction, i.e. it is centered on the true values of the cluster mass profile's parameter. 
The standard deviations are defined assuming the above average uncertainties from real lensing analyses: $\sigma_{r_{200}}/r_{200}=0.1$ and $\sigma_{r_\text{s}}/r_\text{s}=0.3$. As for the correlation, we assume $\rho=0.5$, checking that our results show only negligible effects when varying $\rho$ over a reasonable range of values. 
We thus re-sampled our parameter space over the joint (log) likelihood  $\ln\mathcal{L}_{\text{dyn}}(r_\text{s},r_{200},\mathcal{A}_\infty,\phi_\infty,\mathcal{Q})+\ln P_\text{lens}(r_\text{s},\,r_{200})$.
For our reference analysis, as before, we consider a massive halo with $r_{200}=2.0\,\text{Mpc},\,r_\text{s}=0.3\,\text{Mpc}$ and $\sim 600$ tracers in the \textsc{MG-MAMPOSSt} fit. In order to explore the full parameter range, we have considered the rescaled variables used in e.g. \citet{Terukina:2013eqa,Wilcox:2015kna} 
\begin{equation}\label{eq:scaledvar}
\mathcal{Q}_2=\frac{\mathcal{Q}}{1+\mathcal{Q}},\,\,\,\,\,\,\, \phi_2=1-\exp\left[\frac{-\,\phi_{\infty}}{(M_\text{P}c^2_l 10^{-4})}\right],
\end{equation} 
which run in the interval $[0,1)$.

\begin{figure*}
\centering
\includegraphics[scale=0.55]{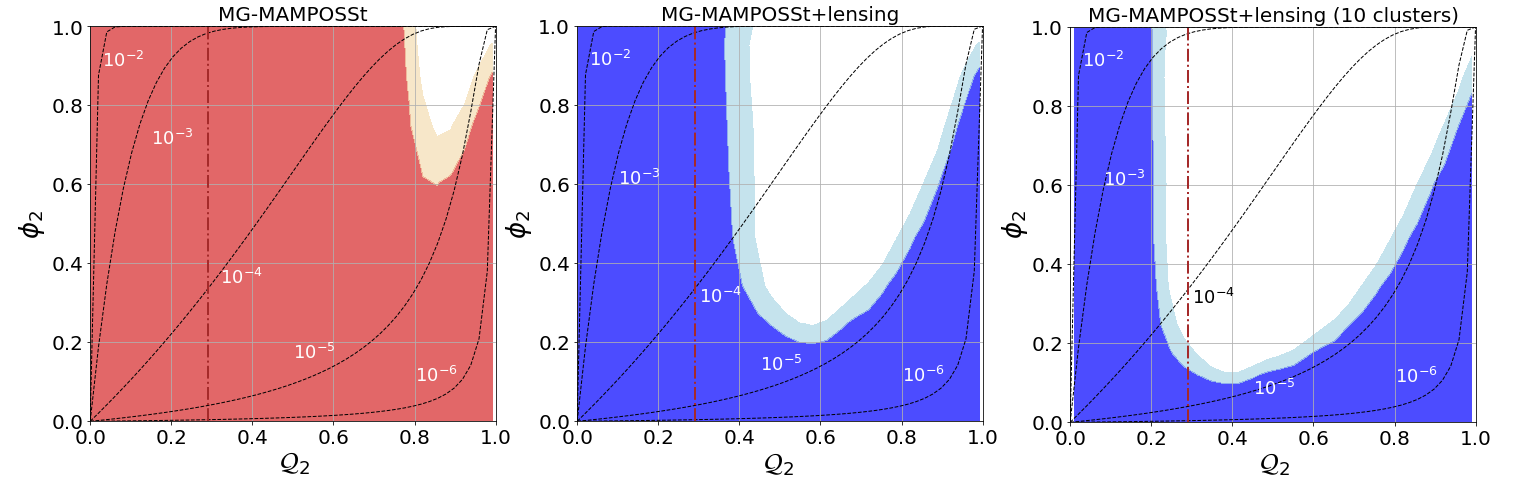}
\caption{\label{fig:cham_ref1} {\it Left and central panels}: results for the MCMC run in CS for one cluster in the sample. The shaded areas  show the allowed regions in the plane of reduced coupling and scalar field,  Eq. \eqref{eq:scaledvar}, $\left(\mathcal{Q}_2=\mathcal{Q}/(1+\mathcal{Q}),\,\,\phi_2=1-\exp\left[-\phi_{\infty}/(M_\text{P}c^2_l 10^{-4})\right] \right)$  at $3\sigma$ (outer region) and $2\sigma$ (innermost region).  {\it Left panel}: results from the \textsc{MG-MAMPOSSt} likelihood. {\it Central panel}: results from the joint lensing+internal kinematics analysis with $\sigma_{r_{200}}/r_{200}=0.1$, $\sigma_{r_\text{s}}/r_\text{s}=0.3,\,\rho=0.5$ in the additional lensing distribution for the same cluster. In both plots 600 tracers are considered in the \textsc{MG-MAMPOSSt} fit. {\it Right panel}:  joint lensing+internal kinematics for the combination of 10 clusters. In this and the following figures we use 
red -light red for \textsc{MG-MAMPOSSt} alone, blue-light blue for the joint lensing+internal kinematics. The black dashed curves indicate lines of constant $\phi/\mathcal{Q}$ while the vertical brown dash-dotted lines correspond to the $f(\mathcal{R})$ value of the coupling $\mathcal{Q}=1/\sqrt{6}$.}
\end{figure*}

The results of the MCMC run for the MG parameters are shown in the left and central plot of Fig. \ref{fig:cham_ref1}. These results are similar to \citet{Terukina:2013eqa,Wilcox:2015kna}, who performed joint lensing and X-ray analyses of 58 clusters and of the massive Coma cluster respectively. Due to the degeneracy between the modified gravity parameters and the mass profile parameters, all values of the coupling factor and of the fields are allowed by galaxy internal kinematics, except for a small region corresponding to $0.8 \lesssim \mathcal{Q}_2 \lesssim 0.9$ and $\phi_2\gtrsim 0.8$ (left plot of Fig. \ref{fig:cham_ref1}). When including the lensing information in the analysis, a considerable fraction of the parameter space can be excluded at $3\,\sigma$, (central plot of Fig. \ref{fig:cham_ref1}). 

For small coupling factors, all scalar field values are still allowed by the data, with the screening mechanism active at small $\phi_{\infty}$ (Eqs. \ref{eq:screenpar1}, \ref{eq:screenpar2}). As the screening radius also depends on the coupling, at large $\mathcal{Q}_2$ the CS becomes very efficient also for large field values. In the intermediate region, however, $\phi_2\to 1$ is no more allowed if $\mathcal{Q}_2\gtrsim 0.4$ as the cluster will be only partially screened and the modification of gravity cannot be entirely suppressed. Already with a single cluster,  the combination of \textsc{MG-MAMPOSSt} with additional lensing distribution provide useful information to investigate this intermediate range of values.

Note that the excluded region found with \textsc{MG-MAMPOSSt}+lensing is shifted towards larger $\mathcal{Q}_2$ w.r.t.  what found by  \citet{Terukina:2013eqa,Wilcox:2015kna}. This is not surprising as X-ray analyses and Jeans equation analyses perceive the effect of gravity and the lack of dynamical equilibrium in different ways, despite their sensitivity to the same gravitational potential. 
%In particular, as mentioned before, the radial squared velocity dispersion $\sigma^2_r(r)$ is affected by the mass profile in the outskirt of the cluster (i.e. the integral of Eq. \eqref{eq:sigmajeans} used in the \textsc{MG-MAMPOSSt} method extends from $r$ to $\infty$). On the other side, the solution of hydrostatic equilibrium for the electron gas pressure at a given radius  is given by (e.g.  \citealt{Wilcox:2015kna}):
%\begin{equation}
%P_e(r)=P_{e,0}+\mu m_\text{p}\int_0^r{n_e(s)\left[-\frac{GM(s)}{s^2}-\frac{\mathcal{Q}}{M_\text{P}}\frac{\text{d}\phi(s)}{\text{d}s}\right]\text{d}s},
%\end{equation}
%where $n_e(r)$ is the electron density profile, $m_\text{p}$ the proton mass $P_{e,0}$ the gas pressure at $r=0$ and $\mu$ the mean molecular weight. Thus, in this case the integral is from $0$ to $r$, i.e. the pressure is determined by the contribution \emph{up to} a given radius.

Our results illustrate that the \textsc{MG-MAMPOSSt} procedure, together with strong+weak lensing analyses, is an interesting method to test CS models at cluster scales -- with present and future observations, the combination of different sets of data and methodologies (i.e. X-ray mass profile determinations) can improve the constraints on the allowed region in the plane $(\mathcal{Q}_2,\phi_2)$ and thus on the viability of those theories at cosmological level.
As for VS models, we study how our results change when combining the information of several clusters, by performing a MCMC sampling of
\begin{equation*}
 \ln\mathcal{L}_\text{joint}=\ln \mathcal{L}_\text{tot}(r_\text{s},r_{200},\mathcal{A}_\infty,\phi_\infty,\mathcal{Q})+N_\text{h}\ln P_\text{lens}(r_\text{s},r_{200}),
\end{equation*}
where $N_\text{h}$ is the number of haloes considered in the stack and the expression of $\mathcal{L}_\text{tot}$ is the same as Eq. \eqref{eq:liketot}. As shown in the right plot of Fig. \ref{fig:cham_ref1}, the analysis of 10 clusters produces an improvement of the constraints for $\mathcal{Q}_2 \gtrsim 0.3$, tightening the allowed region in the parameter space.

\begin{figure*}
\centering
\includegraphics[scale=0.55]{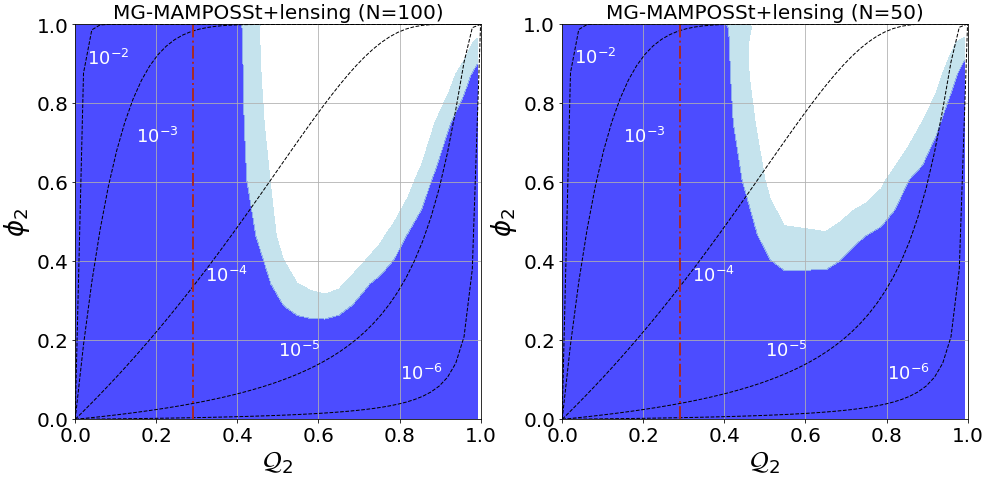}
\caption{\label{fig:lensdyn100} $3\sigma$ (outer region) and $2\sigma$ (innermost region) regions for the joint lensing+internal kinematic analysis of one cluster with $N=100$ (left plot) and $N=50$ (right plot) tracers considered in the \textsc{MG-MAMPOSSt} fit.}
\end{figure*}

Since only few clusters have 600 known redshifts,
we also reduce the number of tracers to $N = 100$, and display the iso-probability contours in the left plot of Fig. \ref{fig:lensdyn100} for the joint lensing+kinematic likelihoods.
While the \textsc{MG-MAMPOSSt} analysis alone is insufficient to provide any bounds in  $(\mathcal{Q}_2,\phi_2)$, the joint lensing+internal kinematics contours are almost identical to the previous analysis with $N=600$. This indicates that the constraining power is mostly related to the additional information provided by lensing on the mass profile parameters $r_{200},r_\text{s}$, confirming what was previously found for linear $f(\mathcal{R})$  (Appendix A of   \citealt{Pizzuti19b}).

Keeping fixed $P_{\text{lens}}(r_\text{s},\,r_{200})$, we generate a new phase space considering only 50 tracers to push further this ideal exercise. As we can see in the right plot of Fig. \ref{fig:lensdyn100}, in this case the weakening of the constraints affects the joint lensing+internal kinematics distribution. 
Although useful information can still be extracted in this limiting situation (e.g. by considering several clusters and suitably combining the member galaxy dynamics with real lensing and X-ray data), it is worth to notice that we are neglecting all the systematic effects. Indeed, such a small number of galaxies for each cluster is insufficient to correctly track the underlying total gravitational potential -- incompleteness of the member galaxies sample  as well as the effect of interlopers (i.e. galaxies in the field of view which do not belong to the cluster) can produce a bias in dynamic mass estimation if not properly taken into account (see  e.g. \citealt{Mamon01}).

\begin{figure*}
\centering
\includegraphics[scale=0.55]{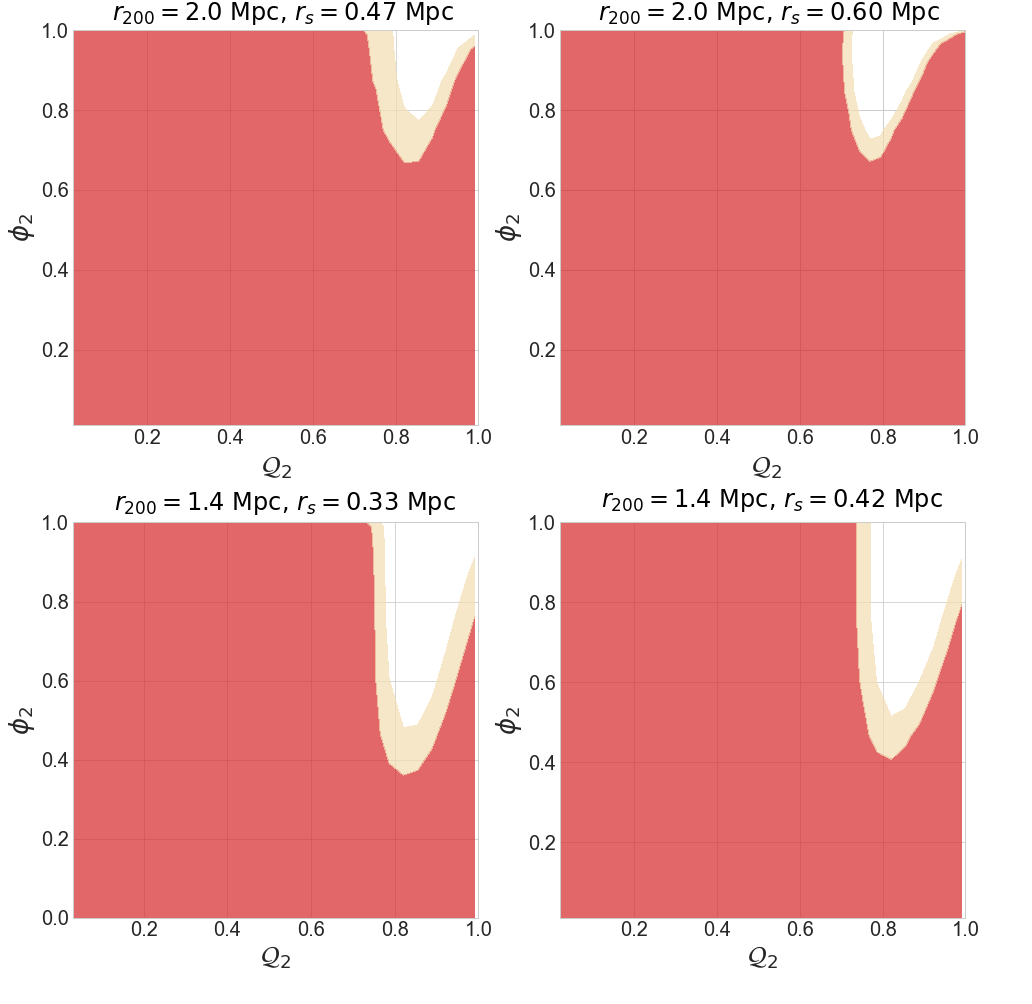}
\caption{\label{fig:varconc} Result of the \textsc{MG-MAMPOSSt} analysis (internal kinematics only) for single haloes with different values of $r_\text{s}$ and $r_{200}$ with respect to the reference case. Color-codes is the same as previous plots. The left upper and left bottom plots refers to a concentration $c=4.2$ obtained changing $r_\text{s}$ and keeping fixed $r_\text{200}$ (up) and vice-versa (bottom). The upper right and bottom right plots show the  result for a concentration parameter $c=3.3$.}% Given the same concentration, the stringent bounds are obtained for the less massive and less concentrated halo.}
\end{figure*}

Whereas for the VS model implemented in \textsc{MG-MAMPOSSt} the new degrees of freedom $Y_1$ and $Y_2$ are scale-independent, this is no more true for CS models. The screening mechanism depends on the structure of the halo density profile, as shown by Eqs. \eqref{eq:screenpar1}, \eqref{eq:screenpar2} -- thus, the analysis of the member galaxy internal kinematics for clusters with different $r_\text{s},\,\,r_{200}$ should produce different constraints. For this reason, we consider four haloes generated with other combinations of $r_\text{s},\,r_{200}$, in agreement with current observations, as discussed in Section \ref{sec:setup}. 
The new phase spaces are in two pairs characterised by concentrations  $c=4.2$ and $c=3.3$ respectively, but obtained by changing $r_\text{s}$ for one halo in the pair and $r_{200}$ for the other halo with respect to the reference case $r_{200}=2.0\,\text{Mpc},\,\,r_\text{s}=0.3\,\text{Mpc},\,\,c=6.67$. We apply again the \textsc{MG-MAMPOSSt} technique on phase spaces with $\sim 600$ particles within $r_{200}$ drawn from those haloes -- we derive the marginalized iso-probability contours for the modified gravity parameters, shown in Fig. \ref{fig:varconc}. As expected, changing both $r_\text{s}$, $r_{200}$ has an impact on the constrained allowed region, with the strongest constraint found for the highest value of the concentration and the smallest virial radius. Note that varying $r_{200}$ produces a larger footprint on the bounds with respect to the change of $r_\text{s}$. This is not surprising, as the screening radius $S$ depends on $\rho_\text{s}r_\text{s}^3\propto r_{200}^3$ (Eq. \ref{eq:screenpar1}).

\subsection{$f(\mathcal{R})$ gravity}
We now restrict to $f(\mathcal{R})$ gravity, a sub-class of chameleon theories where the coupling parameter is fixed to $\mathcal{Q}=1/\sqrt{6}$. Constraining this model translates into bounds on the background value of the scalaron field $|f_{\mathcal{R}0}|$, which mediates the additional fifth force and it is related to $\phi_\infty$ as
\begin{equation}
\phi_\infty=-\sqrt{\frac{3}{2}}\ln(1+f_{\mathcal{R}0})M_\text{P}c^2_l\,,
\end{equation}
according to Eq. \eqref{eq:frcham}.

For our reference analysis, we consider again a phase space of $\sim 600$ particles within $r_{200}$ from an halo with $r_{200}=2.0\,\text{Mpc}$, $r_\text{s}=0.3\,\text{Mpc}$. In the lensing likelihood we adopt the same values of $\sigma_{r_{200}}$, $\sigma_{r_\text{s}}$ and $\rho$ as for our analysis in VS. We then vary the number of haloes considered in the MCMC run, the lensing standard deviations of $r_{200}$ and $r_\text{s}$ and the number of tracers in the internal kinematics analysis.

 When considering a single halo, we found that $\phi_{\infty}$ is unconstrained at $95\%$ C.L. in both cases of $\sim 600$ and $\sim 100$ tracers in projected phase space and for $\sigma_{r_{200}}/r_{200}=0.1,\, 0.07$. This can be seen by looking at the central plot of Figure \ref{fig:cham_ref1}: when $\mathcal{Q}=1/\sqrt{6}$ (brown dash-dotted line) all values of $\phi_2$ are allowed within 2$\sigma$. Moreover, changing the uncertainties $\sigma_{r_\text{s}}$ in the lensing distribution produces a negligible effect on the joint likelihood. This is due to the degeneracy between $r_{200}$ and the chameleon field,confirming what found in the simplified picture of linear $f(\mathcal{R})$ by   \citet{Pizzuti19b}. Indeed, if we fix the coupling constant $\mathcal{Q} = 1/\sqrt{6}$, very large $\phi_\infty$ produces a maximum enhancement of the GR gravitational potential by a factor $1/3$ which can be compensated by slightly decreasing the overall radius of the halo. More in particular, a cluster with a smaller $r_{200}$ in a strong modified $f(\mathcal{R})$ gravity ($\phi_\infty\gtrsim 10^{-4}$) produce the same effect of a cluster with a larger $r_{200}$ in a GR scenario.
 In order to obtain a (weak) bound on $\phi_\infty$, the uncertainties on $r_{200}$ in the lensing Gaussian should be reduced down to $5\%$. In this case we found $\phi_{\infty}\lesssim 9.52\times10^{-5}$ at $2\,\sigma$, corresponding to $|f_{\mathcal{R}0}|\lesssim 7.77\times10^{-5}$.

\begin{figure}
\centering
\includegraphics[width=\columnwidth]{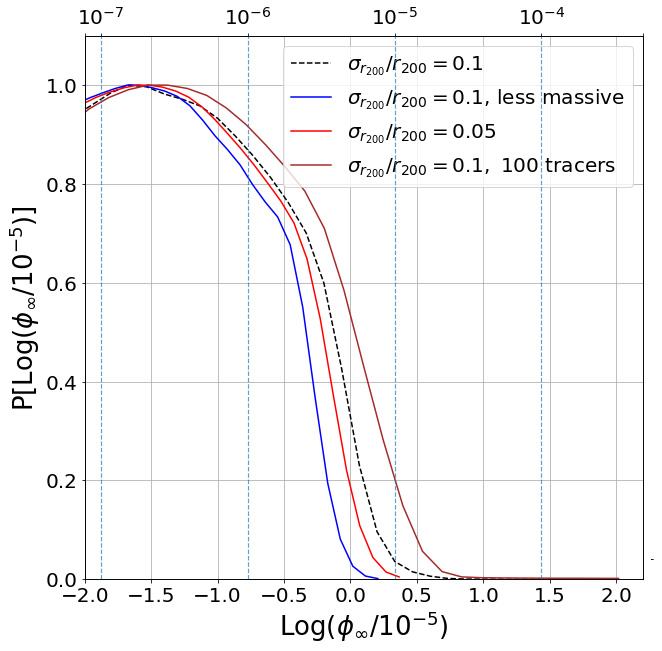}
\caption{\label{fig:combinedfr} Marginalised distribution of the chameleon field in the case of $f(\mathcal{R})$ gravity ($\mathcal{Q}=1/\sqrt{6}$) obtained from the combined, joint lensing+internal kinematics likelihood for 10 clusters. In the lensing Gaussian we consider $\rho=0.5,\,\sigma_{r_\text{s}}=0.3\,r_\text{s}$ and a variable $\sigma_{r_{200}}$. Black dotted line refers to the reference analysis with 600 tracers in \textsc{MG-MAMPOSSt} and $\sigma_{r_{200}}/r_{200}=0.1$. The brown curve shows the case of 100 tracers while the blue distribution is for a phase space of 600 tracers but derived from a cluster with $r_{200}=1.4\,\mpc$ and $r_{s}=0.33\,\mpc$. Finally, the red line refers to $\sigma_{r_{200}}/r_{200}=0.05$. The dashed vertical lines indicate corresponding particular values of $|f_{\mathcal{R}0}|$.}

\end{figure}

As for the general CS model, we perform a stacked analysis to study the constraining power of our method in the ideal situation. In Fig. \ref{fig:combinedfr} we plot the marginalised likelihood obtained by Monte-Carlo sampling over a combined lensing+internal kinematics likelihood of 10 haloes. The black-dashed, blue and red lines indicate different values of $\sigma_{r_{200}}/r_{200}$ in the lensing Gaussian, 10\% (black line) and 5\% (red line). The uncertainties on $r_\text{s}$ are fixed to 30\% and the correlation to $\rho=0.5$. The brown curve is obtained with 100 tracers considered in the \textsc{MG-MAMPOSSt} fit in each of the 10 haloes.

For $\sigma_{r_{200}}= 0.1\,r_{200}$ and $\sigma_{r_{200}}= 0.05\,r_{200}$ the combined distributions are almost identical, indicating that the information which can be obtained with our method for this particular model is already saturated with $\sim 10$ galaxy clusters. This extends the results in Appendix A of   \citet{Pizzuti19b} to the case of chameleon $f(\mathcal{R})$ gravity. For the reference analysis ($\sigma_{r_{200}}/r_{200}=0.1$ in the lensing distribution, black dotted line in of Fig. \ref{fig:combinedfr}) and 10 haloes considered in the combined likelihood we obtain (in terms of $f_{\mathcal{R}0}$):
\begin{eqnarray}
    |f_{\mathcal{R}0}| &\le & 8.26\times 10^{-6} \,\,\,\, 1\sigma,
\\
    |f_{\mathcal{R}0}|&\le & 1.12\times 10^{-5} \,\,\,\, 2\sigma.
\end{eqnarray}
When decreasing the number of tracers down to 100, the constraints become slightly weaker, $|f_{\mathcal{R}0}|\le 3.24\times 10^{-5}$ at $2\sigma$, but still good if compared with the current cosmological constraints on $f(\mathcal{R})$. If the number of haloes in the combined likelihood is extended up to 20, the constraints improve to $|f_{\mathcal{R}0}|\le 1.79\times 10^{-5}$ and $|f_{\mathcal{R}0}|\le 7.11\times 10^{-6}$ at $2\sigma$ for 100 tracers and 600 tracers in the \textsc{MG-MAMPOSSt} fit, respectively. The results obtained varying the number of clusters considered in the analysis are listed in the last two columns of Table \ref{tab:Y}.

As mentioned before, the bounds can be substantially tightened by considering a less efficient screening mechanism, i.e. a smaller-size halo. The blue curve in Fig. \ref{fig:combinedfr} is obtained for 10 haloes with $r_{200}=1.4 \,\mpc$ and $r_\text{s}=0.33 \,\mpc$. In this case, the upper limits become
\begin{eqnarray}
    |f_{\mathcal{R}0}|&\le & 3.48\times 10^{-6} \,\,\,\, 1\sigma,
\\
    |f_{\mathcal{R}0}|&\le & 5.40\times 10^{-6} \,\,\,\, 2\sigma.
\end{eqnarray}
The constraints further strengthen to $|f_{\mathcal{R}0}|\le 3.56\times 10^{-6}$ at $2\sigma$ in the case of 20 haloes.
The bounds achieved by this simple exercise with a reasonable number of clusters are of the same order of magnitude of the results obtained from other cosmological probes (see e.g.  \citealt{Wilcox:2015kna,Cataneo:2016iav, Jana_2019}) for the specific class of $f(\mathcal{R})$ models. As already mentioned, it will be interesting to further evaluate the effect of X-ray analyses which could provide additional information to the dynamical parameters. We point out again that our results are obtained neglecting all systematics, as a proof of concept of the statistical power of the \textsc{MG-MAMPOSSt} method.

\section{Summary and Conclusions \label{sec:Summary}}
We have presented a new extension of the \textsc{MG-MAMPOSSt} code, aimed at constraining modified gravity models with the internal kinematics of the member galaxies in clusters. In particular, we have implemented two parametrisations for the Newtonian potential $\Phi$, which governs the dynamics of member galaxies, based on Vainshtein screening and generic chameleon screening, both with NFW mass density profiles. These models cover the range of viable scalar-tensor theories after the recent measurement of the speed of gravitational waves. The code is based on the original \textsc{MAMPOSSt} method, developed by   \citet{Mamon01}, which applies a Maximum Likelihood approach by solving the spherical Jeans equation in the projected phase space of member galaxies $(R,v_z)$. 
As shown in Section \ref{sec:Codes}, \textsc{MG-MAMPOSSt} can be used to fit together the mass and number density profile parameters, the velocity anisotropy parameter plus the additional degrees of freedom related to the modified gravity theory under study.

%The code has been equipped with a simple MCMC module based on a Metropolis-Hastings algorithm for an efficient exploration of the parameter space of the theories. A run takes few hours to compute the likelihood for $\sim 10^5$ sampled values of the parameters for the case of all the implemented models. 

Under the assumptions of dynamical relaxation and spherical symmetry, we have illustrated the code's capabilities for a set of synthetic isolated dark matter haloes, generated by using the \textsc{ClusterGEN} code of   \citet{Pizzuti:2019wte}. In this idealised setup, we have investigated the statistical degeneracy between the model parameters for VS and CS models that could obscure future realistic constraints in this context. We have further produced forecasts on those parameters in view of current and future observational surveys. The results are summarized as follows:
\\

 In the case of beyond Horndeski gravity we found that the resulting likelihood computed with \textsc{MG-MAMPOSSt}, based on the internal kinematics of member galaxies, exhibits a considerable degeneracy between the NFW parameters and the coupling constant $Y_1$, which cannot be removed even when considering stacked haloes (see Figure \ref{fig:horncomb5} and the second column of Table \ref{tab:Y}). Additional information is required to break the degeneracy and constrain this class of modified gravity models.

Since for this family of models lensing is also affected as compared with GR, we simulated additional weak lensing information on the cluster's mass by considering a mock tangential shear profile with reliable uncertainties. We showed that the forecasted constraints obtained on the fifth force coupling $Y_1$ (associated to the Newtonian potential $\Phi$) are of order unity, significantly weaker to other astrophysical probes. In particular, by considering 20 haloes' phase spaces with 600 tracers in the \textsc{MG-MAMPOSSt} fit, we obtain $Y_1=0.08^{+0.77}_{-0.34}$. When reducing the number of galaxies in each phase space to 100 the constraints weaken to $Y_1 \lesssim 1.02$.   However, as we further showed, the joint internal kinematics+lensing analysis can be used to place tight constraints on the parameter $Y_2$ which appears in the relativistic potential $\Psi$ and it has a stronger impact on the observed mass profile with respect to $Y_1$. In this case, from the aforementioned analysis we found $Y_2=0.01^{+0.09}_{-0.08}$ and $Y_2=0.01^{+0.16}_{0.14}$ for 600 and 100 tracers in the fit respectively. The results are summarised in Figure \ref{fig:lensing1} and in columns three to eight in Table \ref{tab:Y}. It is important to point out that galaxy cluster mass profiles provide a complementary, independent approach to test beyond Horndeski theories w.r.t astrophysical analyses, which relies on different physics and scales.
\\

For CS, we have considered the general case with two free parameters, the fifth-force coupling strength $\mathcal{Q}$ and the background value of the scalar field $\phi_\infty$. The \textsc{MG-MAMPOSSt} code together with additional (lensing) information, provides useful insights on the allowed region of the parameter space, in terms of the re-scaled parameters $\mathcal{Q}_2=\mathcal{Q}/(1+\mathcal{Q}),\,\,\,\phi_2=1-\exp\left[-\phi_{\infty}/(M_\text{P}c^2_l 10^{-4})\right]$. We found that, given the dependence of the velocity dispersion on the gravitational potential, the module implemented in \textsc{MG-MAMPOSSt} can be used to get complementary information to X-ray mass reconstructions, even if they are both sensitive to the Newtonian potential $\Phi$. 

In this class of models, lensing determinations are not affected by the fifth force, thus they can be used as a prior information on the cluster parameters $r_\text{s}\,,r_{200}$. Those priors have a strong impact on the constraining power of the method. Indeed, we found that decreasing the number of tracers from 600 to 100 in the \textsc{MG-MAMPOSSt} fit but keeping the same simulated lensing distribution produces a negligible effect on the two dimensional iso-probability contours in the plane ($\mathcal{Q}_2$, $\phi_2$), as shown in Figure \ref{fig:lensdyn100}. 

For our reference analysis of a phase space with 600 galaxies and $\rho=0.5,\,\sigma_{r_\text{s}}=0.3\,r_\text{s}$, $\sigma_{r_{200}}=0.1\,r_{200}$ in the lensing (Gaussian) distribution we found that the region corresponding to $0.45\lesssim\mathcal{Q}_2\lesssim 0.7$ and $\phi_2\gtrsim 0.6$ is excluded at more than $3\sigma$. The combined likelihood of 10 haloes increases the range of the prohibited area up to $\phi_2\gtrsim 0.3$ for values of $\mathcal{Q}_2$ between $0.3$ and $0.6$.

The CS depends on the matter density perturbations, producing a stronger damping of the fifth force in more massive clusters. In Figure \ref{fig:varconc} we considered four different NFW haloes characterised by distinct values of $r_\text{s}$ and $r_{200}$. We confirmed that smaller-size haloes provide better constraints due to a less-efficient screening mechanism. For clusters with $r_{200}=1.4 \,\mpc$, the \textsc{MG-MAMPOSSt} fit alone can exclude at $3\sigma$ the part of the parameter space where  $\phi_2\gtrsim 0.7$, $\mathcal{Q}_2 \gtrsim 0.7$.
    
As a final step, we have focused on the specific sub-class of chameleon $f(\mathcal{R})$ gravity (i.e. $\mathcal{Q}=1/\sqrt{6}$, $\phi_\infty=-(3/2)^{1/2}\ln(1+f_{\mathcal{R}0})$ ), in order to forecast the value of the background scalaron field. We obtain, for the reference case, $|f_{\mathcal{R}0}|\lesssim 1.12\times 10^{-5}$ and $|f_{\mathcal{R}0}|\lesssim 7.11 \times 10^{-6} $ at 95\% C.L. from the joint lensing+internal kinematics analysis of 10 clusters and 20 clusters, respectively. The bounds can be improved when using less massive clusters -- in particular we found $|f_{\mathcal{R}0}|\le 5.40\times 10^{-6}$ and $|f_{\mathcal{R}0}|\le 3.56\times 10^{-6}$ for 10 and 20 haloes with $r_\text{s}=0.33\,\mpc$ and $r_{200}=1.4 \,\mpc$. The results are shown in Figure \ref{fig:combinedfr} and in the last two columns of Table \ref{tab:Y}. 
\\

In conclusion, \textsc{MG-MAMPOSSt} offers a tool to probe GR and viable scalar-tensor theories at the scale of a galaxy clusters. A key aspect that has to be accounted for is the statistical degeneracy between the mass profile parameters and the MG parameters, at least for the models currently implemented in the code. This degeneracy is best
broken by combining the three probes of galaxy clusters: kinematics, X-rays and lensing. Both ground-based (e.g. Vera C. Rubin Observatory) and space telescopes (e.g. Euclid), together with next generation spectrographs and X-ray observatories  (e.g. the forthcoming Advanced Telescope for High-ENergy Astrophysics, Athena) will provide a large amount of high quality imaging and spectroscopic data for several galaxy clusters, allowing for a wide range of applications of our method.

In the future, we aim at extending the parametrisation of the gravitational potential to other mass density profiles, which would allow to investigate the interplay between the new degrees of freedom and the mass density profile parameters. It should be noted that it is not completely clear that the collapse of structures within all modified gravity theories studied here is accurately described by a NFW profile (see e.g \citealt{Corasaniti20} and references therein).
%As a by-product it would allow to test the robustness may better describe the total matter distribution in galaxy clusters in theories of gravity alternative to GR. Indeed the NFW model, despite its wide range of applicability over different scales, could be not the best profile to reproduce the mass distribution of haloes in a modified gravity scenario (see e.g. \citealt{Corasaniti20} and references therein).} 

The results presented here were obtained assuming the ideal conditions required by the limitations of our method. When dealing with real data, several systematics can affect the measurement producing spurious detection of modified gravity. For example, as shown in  \citet{Pizzuti19b}, the lack of dynamical relaxation and deviations from spherical symmetry have a relevant impact on the projected phase space and should be taken into account by adopting suitable selection criteria. The application of the \textsc{MG-MAMPOSSt} method on the observed phase space of relaxed galaxy clusters will be performed in a forthcoming paper.

\section*{Data Availability}
The data that support the findings of this study as well as the code described in the paper are available from the corresponding author, upon reasonable request. The code will be made publicly available in the future on GitHub at the following link https://github.com/Pizzuti92/MG-MAMPOSSt.

\section*{Acknowledgements}
The authors thank the referee Gary Mamon for his constructive criticism and comments which greatly improved the manuscript. LP acknowledges G. Paggi and M. Calabrese for useful comments and discussions. LP is partially supported by a 2019 "Research and Education" grant from Fondazione CRT. The OAVdA is managed by the "Fondazione Cle\'ment Fillietroz-ONLUS", which is supported by the Regional Government of the Aosta Valley, the Town Municipality of Nus and the "Unite\' des Communes valdotaines Mont-E\'milius. I.D.S. has received support by the Czech Science Foundation GA\v{C}R (Project: 21-16583M), as well as through European Structural and Investment Funds and the Czech Ministry of Education, Youth and Sports (Project CoGraDS- CZ.02.1.01/0.0/0.0/15\textunderscore003/0000437) %I.D.S. is funded by European Structural and Investment Funds and the Czech Ministry of Education, Youth and Sports (Project CoGraDS --- CZ.02.1.01/0.0/0.0/15\_003/0000437) and GACR {IDS: \bf to input the project number}. 



\bibliographystyle{mnras}
\bibliography{all-refs} % if your bibtex file is called example.bib

%\begin{thebibliography}{99}
%\end{thebibliography}

%%%%%%%%%%%%%%%%%%%%%%%%%%%%%%%%%%%%%%%%%%%%%%%%%%

%%%%%%%%%%%%%%%%% APPENDICES %%%%%%%%%%%%%%%%%%%%%

%\appendix



%%%%%%%%%%%%%%%%%%%%%%%%%%%%%%%%%%%%%%%%%%%%%%%%%%

% Don't change these lines
\bsp	% typesetting comment
\label{lastpage}
\end{document}